\documentclass[11pt, a4paper]{article}

\usepackage[utf8]{inputenc}
\usepackage[T1]{fontenc}
\usepackage[margin=2.5cm]{geometry}
\usepackage{microtype}

\usepackage{authblk}

\usepackage{amsmath,amssymb,amsfonts}
\usepackage{nicefrac}
\usepackage{siunitx}
\usepackage{xparse}

\usepackage{graphicx}
\graphicspath{{./}}
\usepackage{color,array,booktabs,multirow}
\usepackage[caption=false,font=footnotesize,labelfont=sf,textfont=sf]{subfig}
\usepackage{float}
\usepackage{adjustbox}
\usepackage{algorithm}
\usepackage{algorithmic}

\usepackage{textcomp}
\usepackage{url}
\usepackage{xspace}
\usepackage[final]{changes}

\usepackage[colorlinks,urlcolor=blue,linkcolor=blue,citecolor=blue]{hyperref}
\usepackage{cleveref}

\def\BibTeX{{\rm B\kern-.05em{\sc i\kern-.025em b}\kern-.08em
    T\kern-.1667em\lower.7ex\hbox{E}\kern-.125emX}}

\newcommand{\Tec}[1]{\SI[round-precision=0]{#1}{\nano\meter}}

\NewDocumentCommand{\Qformat}{m m o}{%
    \textnormal{{Q}#1.#2%
    \IfValueT{#3}{.#3}%
    }%
}

\NewDocumentCommand{\Qset}{m m g}{%
  \ensuremath{%
    Q_{#1.#2\IfValueT{#3}{.#3}}%
  }%
}

\newcommand{\Qnmk}{\Qformat{n}{m}[k]\xspace}
\newcommand{\Qnm}{\Qformat{n}{m}\xspace}

\newcommand{\Snm}{\Qset{n}{m}\xspace}
\newcommand{\Snmk}{\Qset{n}{m}{k}\xspace}

\DeclareSIUnit\BGunit{\milli\gram\per\deci\liter}
\DeclareSIUnit\Eunit{\joule}
\DeclareSIUnit\mJ{\micro\joule}
\DeclareSIUnit\mAh{\mA\hour}

\begin{document}

\title{Multiplier-less Neural Network Inference for Low-Power Embedded LSTMs\thanks{This work is funded by national funds of the Spanish Ministry of Science, Innovation, and Universities and the Agencia Española de Investigación, through grants PDC2025-165077-I00 MICIU/AEI and PID2024-158129OB-I00 MICIU/AEI/10.13039/501100011033 and by FEDER, UE.}}

\author{Felix Tena\thanks{Corresponding author: \texttt{feltena@ucm.es}}}
\author{Oscar Garnica}
\author{Juan Lanchares Dávila}
\author{J. Ignacio Hidalgo}

\affil{Department of Computer Architecture and Automation, Universidad Complutense de Madrid, Spain}

\date{}

\maketitle

\begin{abstract}
Constrained-Hamming Quantization is a fixed-point quantization framework designed to reduce energy consumption in power-constrained edge devices, particularly for neural networks dominated by multiply-accumulate operations. Constrained-Hamming Quantization bounds the number of active bits within the fixed-point representation of neural network weights to a predefined maximum, while retaining the original word width and fixed-point scaling, preserving numerical granularity near zero and preventing small-magnitude coefficients from collapsing to zero. This active-bit bound enables a positional encoding strategy in which the locations of the active bits are stored explicitly. To exploit the resulting representation efficiently in hardware, this work introduces a dedicated arithmetic architecture that replaces conventional multipliers with a shift-and-add arithmetic structure, lowering switching activity and dynamic power dissipation. To facilitate deployment, we also introduce an automated RTL design flow that trains, quantizes, and maps Constrained-Hamming Quantization models directly onto FPGAs. While the methodology is applicable across diverse network architectures, the proposed approach is evaluated on embedded LSTM models through both hardware-level and system-level experiments, including comparisons against a conventional hardware implementation and a representative mobile edge device. Experimental results show that the proposed arithmetic architecture achieves a \SI{25}{\percent} energy reduction compared with the baseline hardware implementation, maintaining this efficiency advantage across varying model complexities while ensuring the quantization error satisfies the target requirements.
\end{abstract}



\section{Introduction}
\label{sec:intro}

Wearable and other energy-constrained edge devices have emerged as key enablers of real-time data collection and localized processing across various applications, including industrial sensing, lifestyle tracking, and human-computer interaction. Their compact form factor, sensor integration, and connectivity capabilities make them ideal for continuous monitoring of physiological and environmental parameters. While many of these devices are designed for general consumer use, operating under strict battery constraints makes energy efficiency their primary design challenge. This context has opened new opportunities for integrating advanced AI-based models directly into energy-constrained nodes, establishing architectural improvements applicable to both specialized biomedical systems and general battery-powered wearables \cite{kazanskiy2024}.

With these battery-powered nodes integrated into daily life, a large volume of data is acquired, providing real-time insights into environmental and operational states. Consequently, there has been a significant increase in the development and deployment of AI algorithms specifically designed to harness this wealth of data directly on the capturing nodes. Among AI techniques, neural networks (NNs) have become the most popular machine learning approach, thanks to the success of models created by large companies \cite{deng2020}. NNs excel at processing complex signals without relying on hand-crafted features \cite{hamet2017}. These algorithms automatically analyze intricate patterns within the collected data, offering predictive analytics and anomaly detection among others. The different NNs can use diverse data, including physiological measurements, to learn how to solve specific problems that are too complex to be solved manually.

Among the different existing NNs, long short-term memory (LSTM) NNs, stand out as one of the most suitable for processing sequential inputs on portable platforms. Their popularity stems from their structural simplicity and their ability to process time-series data, recording changes in variables of interest over time, which is fundamental for continuous tracking applications . Even though LSTMs are moderately simple compared to other NNs capable of processing temporal data, such as transformers, they require complex arithmetic operations. For this reason, energy consumption becomes the primary constraint in portable devices, particularly for applications that demand uninterrupted operation in power-constrained environments \cite{zhu2022}.

To address energy consumption issues, cloud computing has become a well-known solution, as it reduces the need for local infrastructure and alleviates local power demands \cite{moha2018}. Major cloud service providers such as Google, AWS, and Azure, which operate large-scale data centers, make this possible. These centralized infrastructures enable the sharing of hardware components, thereby minimizing redundancy. Furthermore, their vast resource capacity enables them to optimize the computational operations required for training and deploying AI algorithms efficiently \cite{dong2016}.

One of the key components that enhances the energy efficiency of cloud computing is the use of systolic array-based matrix multiplication units (MXUs). Systolic design is a hardware architecture that allows data to flow through a network of processing elements. This approach reduces energy consumption by enabling values to flow directly across multiple operations, avoiding repeated storage in registers, which is one of the primary energy costs in traditional computing architectures. By enabling data to be passed directly between adjacent processing units, systolic arrays significantly increase throughput and reduce latency, making them particularly well-suited for the high-volume matrix operations required in machine learning workloads \cite{sato2017}.

However, cloud computing has three main problems related to its intrinsic characteristics. First, the distance between the edge device and the cloud increases both latency and energy consumption \cite{shankar2001}. In addition to the energy used for inference, the energy required to transmit data from the edge to the cloud must also be considered, potentially increasing total consumption by an additional \SIrange[round-precision=0]{10}{50}{\percent} per inference \cite{Desislavov2023, Schwartz2020}. Second, security and privacy concerns associated with sharing data with the network may pose a potential problem when working with sensitive data, such as physiological measurements . Third, continuous operations that cannot be stopped due to a loss of connectivity that halts the model \cite{bavli2024}.

To address the challenges mentioned above, edge computing has been gaining significant traction in recent years as a solution. The fundamental concept of edge computing is to position the computational hub outside the network, thereby separating it from the cloud. This strategy avoids unwanted latency, enhances real-time experience, strengthens data security, and reduces energy consumption associated with communication processes. Hence, NNs, known for the quick analysis of voluminous data, have aroused interest in their implementation at the edge \cite{deng2020}. Consequently, once NNs are trained, they can be deployed at the edge by applying several optimization techniques. 

At first glance, smartphones may appear to be ideal platforms for implementing AI at the edge. The execution of NNs on smartphones eliminates the latency associated with cloud connectivity, enhances the user experience, and reduces the energy consumption required for continuous data transmission. In the current digital age, smartphones are the primary devices for collecting, processing, and sharing personal data. By retaining this data at the edge, data privacy is enhanced, enabling the operation of NNs offline when internet connectivity is intermittent or unavailable \cite{xun2020}.

Despite the performance of smartphones nowadays, they are inherently limited by their constrained computing power and battery life. Furthermore, the generic hardware of smartphones results in suboptimal processing speed and energy consumption. Consequently, with these limitations, an NN could potentially monopolize the computational resources and drain the battery, rendering the smartphone incapable of performing other functions. Therefore, while smartphones offer certain advantages, their limitations must be carefully considered in the context of edge AI implementation.

In response to the limitations inherent in smartphones, Field-Programmable Gate Array (FPGA) devices, which are configurable integrated circuits, present a viable alternative. FPGA-based accelerators \cite{Nechi2023} can reduce the energy consumption of NN inference operations compared to generalist hardware, utilizing techniques designed to fine-tune the trained NN for low-energy and fast operations \cite{Shawahna2019}. In addition, FPGAs are suitable for prototyping NN hardware accelerators for energy-constrained edge devices, offering flexibility, moderate cost, and high performance. Their reprogrammable nature makes them an optimal choice for implementing LSTMs and NNs, in general, at the edge. The flexibility of FPGAs allows for replacing one NN with another that has shown superior performance. Moreover, they can be fine-tuned to create a personalized model for each user, which is highly valuable in domains requiring continuous adaptation, such as personalized healthcare. This adaptability and potential for personalization make FPGAs a highly effective tool in edge computing. FPGAs also serve as a fair comparison point against smartphones, the most accessible Commercial Off-The-Shelf platforms for running NNs, as they require no hardware modifications and provide mature development environments \cite{tena2023}.

Nevertheless, deploying LSTM NNs on energy-constrained edge devices remains a significant hardware challenge, primarily due to the power consumption and area overhead associated with standard MXUs. While FPGAs provide a low-power alternative to general-purpose processors, conventional hardware implementations still rely on dense arithmetic multipliers to compute NN activations and recurrent operations. 

To address this limitation, this work introduces a constrained-Hamming quantization (CHQ) framework specifically designed for energy-efficient NN inference. First, we propose a constrained active-bit fixed-point representation that limits the number of non-zero terms within the quantized weights while preserving sufficient numerical precision for time-series prediction tasks. Second, we analyze different binary representations and reduction mechanisms, demonstrating that sign-magnitude(SM) encoding provides a more favorable quantization error distribution for aggressive active-bit reduction. Third, we introduce a positional encoding strategy that stores the locations of active bits directly in memory, eliminating the need for runtime bit-search logic. Finally, we present a specialized multiplier-less arithmetic architecture that exploits this encoded representation to replace standard matrix multiplications with lightweight shift-and-add operations. Together, these contributions establish a representation-driven arithmetic framework that significantly reduces computational overhead and energy consumption across resource-constrained edge platforms.

To maximize the efficiency of the CHQ representation, the hardware architecture is explicitly designed around systolic array principles to optimize data flow. Rather than mapping the NN to a generic computation engine, the process is decomposed into a sequence of interconnected processing nodes. Each node integrates the CHQ-driven logic to perform partial operations on streaming input data, enabling a fully pipelined, spatially distributed execution. This combination of CHQ arithmetic and systolic data flow enables highly specialized hardware resources. Furthermore, to facilitate the practical deployment of this complex architecture, we provide an automated design flow that trains, quantizes, and outputs a complete Register Transfer Level (RTL) implementation of the CHQ-based model, ready for immediate FPGA integration.

Moreover, the energy efficiency of the proposed hardware implementation is comprehensively evaluated. While customized edge hardware inherently reduces power consumption compared to general-purpose devices, it is necessary to isolate the specific energy savings achieved by our proposed arithmetic. Therefore, the energy consumption of the constrained-Hamming arithmetic architecture deployed on an FPGA is compared against two distinct baselines: a state-of-the-art smartphone and our previous multiplier-based hardware LSTM implementation \cite{tena2023}. This dual comparison allows us to study how energy consumption scales with NN depth and to demonstrate the specific advantages of CHQ arithmetic in real-world scenarios.

The remainder of this article is organized as follows. In \Cref{sec:rw}, we explore related works to illustrate the motivation behind our architectural choices. \Cref{sec:framework} details the proposed CHQ framework, providing the mathematical basis of our quantization proposal, the hardware architecture, and the automated design flow. \Cref{sec:results} establishes the foundational experimental setup, defining the dataset, the baseline NN architecture, and the general methodology applied across all subsequent evaluations. \Cref{sec:representation_analysis} analyzes the CHQ representation and transformation design space to identify the optimal quantization configuration. \Cref{sec:multiplier_implementation} assesses the hardware performance of the arithmetic architecture. \Cref{sec:system_level} evaluates the predictive accuracy, energy consumption, and scalability of the complete NN implementation. Finally, \Cref{sec:conclusion} summarizes the findings of this study.

\section{Related work} \label{sec:rw}

This section reviews the integration of AI into wearable devices, with a focus on medical applications, as this domain represents an area with a potentially high demand to implement our proposed solution, and it is one of the most demanding fields regarding precision.We first discuss the enabling technologies behind AI-powered wearables, followed by AI's role in healthcare tasks like signal analysis and disease prediction, particularly using LSTM models for time-series data. We then explore recent efforts in edge AI for wearables and examine current FPGA-based solutions for NN deployment, framing the motivation for our proposed CHQ architecture.

\subsection{Wearable Edge AI}

Recent advances in sensor miniaturization, low-power electronics, wireless communication, and embedded processing have made it possible to integrate AI into wearable devices, enabling real-time, on-body inference. While these applications extend to sports performance \cite{david2023} and environmental health monitoring \cite{minaee2022}, their impact is particularly notable in healthcare. AI-enhanced medical wearables support continuous physiological monitoring, early disease detection, and personalized interventions, improving healthcare accessibility and outcomes \cite{wang2025}.

While AI algorithms applied to the medical field have demonstrated strong performance in medical image analysis, treatment assistance, and extracting information \cite{amisha2019, Mayo2017, Esteva2017, Lakhani2017, garnica2021}, our focus is on the analysis of time-series data, specifically physiological measurements acquired over time. LSTMs have emerged as effective tools for addressing the challenges associated with such data. Among others, LSTMs have shown high performance in predicting blood glucose (BG) levels for Type 1 Diabetes Mellitus \cite{tena2021}, heartbeat classification \cite{hiri2021}, and detecting emotions using physiological signals \cite{awais2021}. These examples summarize the potential of LSTMs in healthcare, highlighting their ability to leverage time-series data for improved disease prediction, diagnosis, and treatment while preserving the complexity of AI algorithms, which is crucial for edge AI applications.

A comprehensive review of edge AI applications in healthcare systems was conducted in \cite{badidi2023}, which explores the deployment and execution of different AI techniques implemented as matrix operations at the edge, and reviews the role of Edge AI in early health prediction. Additionally, another extensive review delves into the influence of AI on personalized health monitoring \cite{olyan2024}. Specific case studies include a system that gathers cardiovascular data and employs machine learning algorithms in real time using data from smartwatches \cite{eko2022}, an approach to enhance the use of wearable monitoring systems to classify electrodermal activities for stress detection \cite{almadhor2023}, and an integration of wearable devices and edge AI technologies tailored to prevent chronic heart failure \cite{shumba2023}. These works highlight the capacity of edge intelligence to integrate wearable devices into infrastructures for patient diagnosis and management while preserving data privacy.

Consequently, the wide variety of medical applications, demonstrates the extensive potential of edge intelligence in healthcare. As the demand for continuous patient monitoring expands across different medical disciplines, the need to deploy complex NNs on power-constrained devices becomes increasingly relevant. This growing landscape of wearable medical solutions underscores the necessity for optimized hardware architectures capable of sustaining continuous physiological analysis directly at the edge.

\subsection{Weight Quantization Strategies}

To enable this edge integration, deploying complex time-series models like LSTMs onto power-constrained devices requires aggressive mathematical compression. Weight quantization strategies have emerged as a primary method for deploying deep learning models on resource-limited hardware, such as FPGAs, by significantly reducing storage requirements and computational complexity. These strategies focus on representing NN weights with low precision, moving away from traditional 32-bit floating-point parameters that demand extensive memory and high-performance processing units. By compressing the parameter space, quantization enables the deployment of complex models on devices with limited computational resources and power constraints.

The most aggressive form of these strategies is binarization, which uses 1-bit quantization, with weights restricted to two possible values, typically $-1$ or $+1$. This approach allows for the replacement of heavy matrix multiplications with lightweight bitwise XNOR and Bitcount operations. While this extreme reduction in precision yields up to 32x memory savings and power efficiency, it introduces challenges, including severe information loss and optimization difficulties arising from the non-differentiability of the binarization functions. To address these drawbacks, various optimization-based algorithms have been developed, ranging from minimizing quantization error via scaling factors to employing specialized training techniques such as the straight-through estimator \cite{qin2020}.

However, these extreme quantization techniques are primarily effective for large-scale NNs. Such architectures, often containing millions of parameters, exhibit significant structural redundancy, which enables them to maintain performance even after binarization. In contrast, small-scale models,  lack this inherent redundancy. In these constrained architectures, each individual parameter becomes critical for capturing temporal dependencies; consequently, the information loss induced by binarization often leads to a catastrophic collapse in the model’s predictive accuracy.

To overcome the accuracy degradation of binarization, fixed-point quantization is widely accepted as a more reliable alternative for hardware deployment. Unlike binary or strictly integer formats, fixed-point representations retain the fractional precision that most NNs require to properly capture subtle data variations, successfully avoiding the massive hardware overhead of full floating-point arithmetic. This necessary balance between predictive accuracy and hardware efficiency was the cornerstone of our previous work \cite{tena2023}. In that study, we demonstrated that quantizing the values using an optimized fixed-point format (specifically, \Qformat{3}{7}) minimized logic block usage and latency without compromising clinical accuracy. However, despite these memory and speed optimizations, fixed-point implementations fundamentally continue to rely on hardware multipliers to compute the operations, leaving a significant energy bottleneck in ultra-low-power devices \cite{lin2016}.

To bypass the high energy consumption of fixed-point multipliers, the literature has extensively explored Power-of-Two (PoT) quantization strategies. In these schemes, weights are constrained to exact powers of two ($2^n$), allowing the hardware to replace complex digital multipliers with simple bit-shift operations. However, restricting weights to a single power of two imposes a rigid, logarithmic distribution with increasingly large gaps between representable values. This non-uniform quantization introduces substantial errors that degrade the model's accuracy. In small, highly sensitive architectures, every parameter is critical, and this magnitude of precision loss worsen the NN's predictive capacity. Consequently, while PoT quantization excels in arithmetic simplicity, it fails to meet accuracy demands, highlighting the need for a numerical representation that provides finer resolution without reintroducing standard multipliers \cite{gholami2021}.

\subsection{FPGA-based NN Accelerators}

Translating the theoretical benefits of weight quantization into actual energy savings requires a suitable hardware substrate. As mentioned in \Cref{sec:intro}, FPGAs are well-suited for implementing NNs in wearable medical devices. The current landscape of FPGA-based AI accelerators spans both commercial vendor platforms and custom academic architectures, each introducing distinct strategies for hardware design and arithmetic execution. Within the commercial sector, these platforms are generally divided into two main categories based on their target applications: high-performance computing and low-power edge deployment.

Major FPGA vendors such as AMD and Intel primarily focus on high-performance AI acceleration by integrating specialized hardware components into their architectures. AMD utilizes a hybrid approach in its Versal family, combining AI Engines (AIEs), functioning as programmable processors, with programmable logic to achieve multi-level parallelism. Similarly, Intel employs hardened AI Tensor Blocks based on systolic arrays of multipliers and adders to accelerate matrix operations with ASIC-like efficiency. Both vendors rely on sophisticated software ecosystems, such as Vitis AI and OpenVINO, to manage model quantization and high-level synthesis (HLS). However, these solutions prioritize computational throughput and architectural generalization, making them less suitable for the stringent energy-efficiency requirements of continuous-monitoring wearable medical devices.

In the low-power domain, Microchip and Lattice Semiconductor provide FPGA architectures specifically tailored for edge AI applications with constrained energy budgets. Microchip’s PolarFire family leverages a RISC-V subsystem integrated with programmable logic and standard digital multipliers (DSP) resources to execute signal-processing tasks within a power budget of a few watts. Similarly, Lattice focuses on small-form-factor devices, utilizing the sensAI stack to map AI workloads onto power-efficient fabrics through soft-core processors and pre-optimized IP cores. While these vendors offer tools for model compression and quantization to fit restricted environments, their implementations still rely on general-purpose DSP slices and soft-core architectures to maintain broad applicability. Consequently, they do not fully exploit the potential for extreme energy savings that can be achieved through custom, multiplier-less arithmetic specialized for specific NN topologies.

To overcome this reliance on standard DSP blocks, the academic community has proposed custom hardware architectures tailored to the specific quantization strategies previously discussed. For instance, hardware accelerators designed for extreme quantization, such as BNNs, physically replace standard arithmetic blocks with simple XNOR logic gates \cite{pappalardo2022}. Similarly, custom architectures leveraging PoT quantization implement shift-and-add mechanisms to eliminate DSP blocks entirely \cite{chmiel2021, poli2022}. While these highly specialized designs maximize energy efficiency by directly mapping the quantized arithmetic to custom logic, they inherit the accuracy limitations of their underlying numerical representations, rendering them unsuitable for precise clinical predictions.

In summary, while current FPGA architectures range from high-performance tensor blocks to low-power edge fabrics, they share a common approach to implementing NNs. Rather than mapping the neural cells as they appear in architectural descriptions, such as following a complete systolic design, these platforms abstract computation to its core linear algebra primitives. This abstraction results in a reliance on general-purpose multipliers, typically DSP slices, for vector-matrix operations. Even in power-efficient devices, these arithmetic units consume a considerable portion of the energy budget due to their hardware-agnostic design.

Up to this point, our analysis has established the critical role of edge AI in medical wearable devices and the effectiveness of LSTMs for processing time-series physiological data. To deploy these models within strict power budgets, we have examined the prominent weight quantization strategies and the hardware approaches offered by both commercial FPGA platforms and custom academic architectures. However, this review reveals that existing methods either cause substantial degradation in clinical accuracy or, when sufficient precision is maintained, fail to eliminate energy-intensive arithmetic operations. Consequently, there is a distinct need for novel, arithmetic-level hardware architectures specifically optimized to meet both the power and precision constraints of wearable medical devices.

\subsection{Motivation and Research Objectives}

In contrast to the generalized approaches discussed in the literature, developing an AI-based wearable medical device requires optimizing energy consumption. While the market offers AI accelerator hardware and generic deployment toolchains, these solutions are generally designed for processing speed, throughput, and architectural flexibility. They typically rely on standard MXUs structured to accommodate multiple NN topologies. Consequently, these architectures tend to exhibit higher power consumption and larger form factors, which pose challenges for the physical and energy constraints of battery-constrained devices. To address this gap, we propose a specialized hardware architecture that adapts systolic design principles through a dual optimization strategy: architectural and arithmetic.

At the architectural level, rather than mapping NN operations onto uniform MXUs, our approach decomposes the NN into specialized computational nodes. Each node is tailored for specific partial computations, avoiding the need for intermediate storage in registers. At the arithmetic level, we address the power consumption bottleneck by replacing DSPs within these nodes with the proposed multiplier-less architecture. By combining specialized, register-free data routing with the CHQ, our design increases circuit specificity, improving energy efficiency while maintaining diagnostic accuracy.

{Although the proposed methodology is widely applicable to a broad class of inference accelerators, an LSTM network for blood glucose prediction was utilized as a demanding testbed to validate the capabilities of this architecture. While newer models, such as transformers, have shown effectiveness in general time-series analysis, they typically require more parameters and computationally intensive attention mechanisms. Furthermore, transformers depend on large datasets for training, which are not always available in personalized clinical environments. In contrast, LSTMs provide a practical balance between predictive capacity and computational cost. Their recurrent architecture maintains consistent performance while adhering to the low-power constraints, making them a suitable benchmark to evaluate the energy efficiency of our CHQ proposal.

\section{Constrained-Hamming Representation and Arithmetic \label{sec:framework}}

In contrast to the generalized approaches discussed in the literature, developing AI-based edge devices requires optimizing energy consumption from the algorithmic level down to the physical implementation. Standard AI accelerator hardware and generic deployment toolchains are generally designed for processing speed and architectural flexibility, relying on standard MXUs structured to accommodate multiple NN topologies. Consequently, these architectures tend to exhibit higher power consumption and larger form factors.To address this gap, we propose a novel representation of NN weights followed by a specific data encoding strategy.

\subsection{Constrained-Hamming Quantization}

To eliminate the reliance on standard hardware multipliers, the first step is to optimize the dense dot products involving the static NN weights at the arithmetic level. The literature has explored binarization and PoT formats to simplify arithmetic operations. However, binarization leads to severe information loss, and PoT imposes a rigid logarithmic distribution with large numerical gaps. To bridge the gap between these extremes, we propose a constrained-Hamming representation that limits the number of active bits (non-zero bits) present in the quantized weights. Given a standard \Qnm fixed-point representation, the proposed representation constraints the active-bit cardinality to a maximum value, $k$, defining the proposed \Qnmk format.

Starting from a conventional \Qnm fixed-point representation, the objective of the proposed constrained-Hamming representation is to reduce the number of active bits while preserving the original fixed-point structure and maintaining a close numerical approximation to the original value. Specifically, the transformed representation preserves the same word width and fixed-point scaling as the original \Qnm format, while restricting the number of active bits to at most $k$. The resulting format is denoted as \Qnmk.

Hereafter, \Qnmk denotes the representation format and \Snmk\ the corresponding set of representable values. Formally,

\begin{equation}
\Snmk = \{x \in \Snm \mid H(x) \le k\} \subseteq \Snm
\end{equation}

where $H(x) = \sum_i x_i$ for $x_i \in \{0,1\}$ is the Hamming weight of the binary representation of $x$, i.e., the number of active bits.

Maintaining the original word width preserves the dynamic range of the representation and retains sufficient numerical granularity near zero, preventing small-magnitude coefficients from collapsing to zero due to aggressive coarse quantization. Furthermore, preserving the original fixed-point structure enables the proposed representation to be integrated into existing NN datapaths and fixed-point hardware pipelines without modifying their arithmetic interfaces, memory organization, or operand widths, thereby facilitating hardware reuse and retrofit arithmetic optimization. Consequently, the proposed representation reduces arithmetic complexity while preserving the essential numerical properties required for stable NN inference.

Hence, whenever the number of active bits exceeds the imposed limit $k$, different transformations (reduction operators) can be applied to transform the original \Qnm representation into a constrained-Hamming counterpart \Qnmk. These transformations determine how excess active bits are removed and how the resulting quantization noise introduced during the active-bit reduction process is distributed. In this work, we explore three  transformations: truncation, rounding, and bit-pattern replacement. 

The first transformation is truncation, $T_k : \Snm \rightarrow \Snmk$ where all active bits of $x \in \Snm$ located after the $k$-th active position are directly removed. More specifically, the binary representation is traversed from the most significant bit (MSB) towards the least significant bit (LSB), retaining only the first $k$ active bits encountered during the scan.

The second transformation is rounding, $R_k : \Qset{n}{m} \rightarrow \Qset{n}{m}{k}$ which evaluates the discarded least significant bits and calculates the final value of the $k$ allowed bits based on the discarded least significant bits. Using the same sequence, rounding evaluates the discarded bits and rounds up the last kept bit to improve the final approximation.

Finally, the bit-pattern replacement transformation, $S_\ell: \Qset{n}{m} \rightarrow \Qset{n}{m}$ is a bit-pattern replacement rule that compacts long runs of $r$ active bits with $r \ge \ell$. Hence, any occurrence of a zero followed by $r$ consecutive ones, with $r \ge \ell$, is replaced by a one followed by $r$ zeroes, that is $0\,1^{r} \rightarrow 1\,0^{r}$, iff $r \ge \ell$. $S_\ell$ transformation can be composed with $T_k$ and $R_k$ transformations. However, the composition is not commutative, i.e., $S_\ell\!\circ\!T_k \ne T_k\!\circ\!S_\ell$ and $S_\ell\!\circ\!R_k \ne R_k\!\circ\!S_\ell$

\Cref{fig:transformations} illustrates these transformations. In \Cref{fig:truncation_1}, the transformation $T_3: \Qset{3}{7} \rightarrow \Qset{3}{7}{3}$ discards two active bits with error, $e = x - T_3(x) = \num[round-precision=4]{0,0703125}$, whereas in \Cref{fig:truncation_2} $e=\num[round-precision=4]{0,078125}$. On the other hand, \Crefrange{fig:rounded_1}{fig:rounded_2} present two examples of $R_3$ with $e=\num[round-precision=4]{0,0546875}$ and $e=\num[round-precision=4]{0,078125}$ respectively. Finally, \Crefrange{fig:substituted_1}{fig:substituted_2} present two examples of the $S_2$ transformation.

\begin{figure*}[htp]
    \centering
    \subfloat[$T_3 : \Qset{3}{7} \rightarrow \Qset{3}{7}{3}$]{%
        \includegraphics[width=0.45\linewidth]{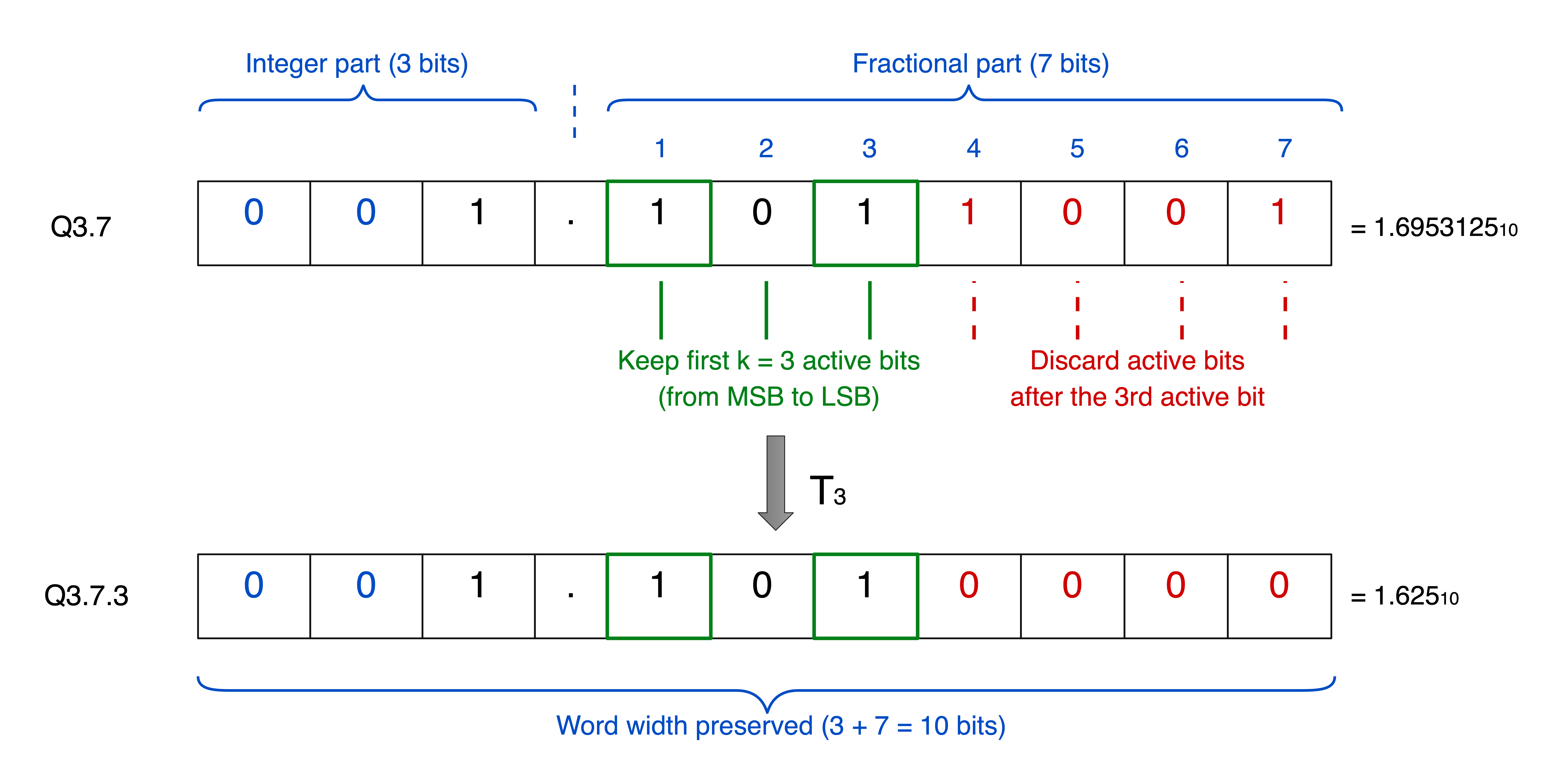}
        \label{fig:truncation_1}
    }
    \subfloat[$T_3 : \Qset{3}{7} \rightarrow \Qset{3}{7}{3}$]{%
        \includegraphics[width=0.45\linewidth]{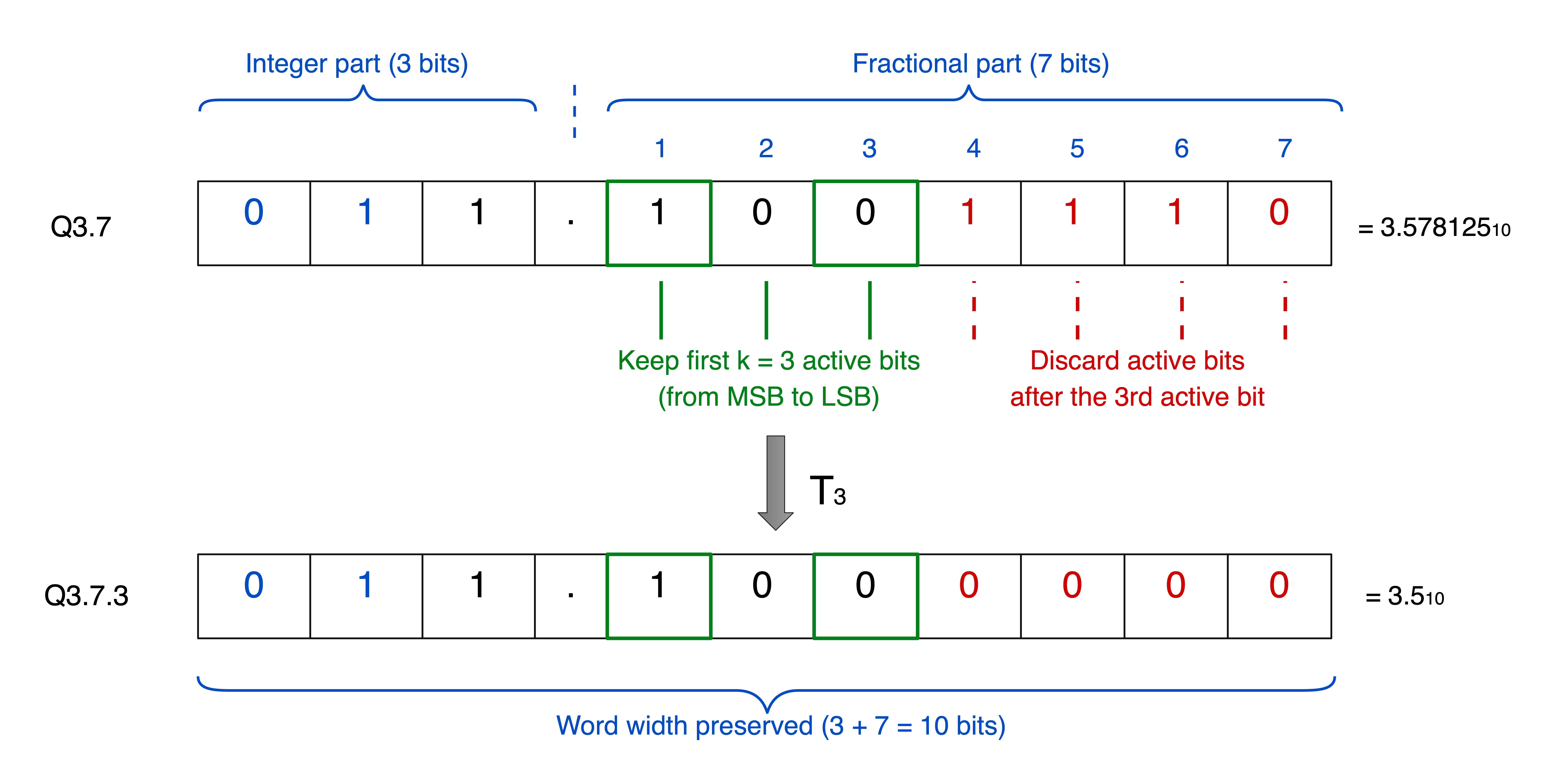}
        \label{fig:truncation_2}
    }\\
    \subfloat[$R_3 : \Qset{3}{7} \rightarrow \Qset{3}{7}{3}$]{%
        \includegraphics[width=0.45\linewidth]{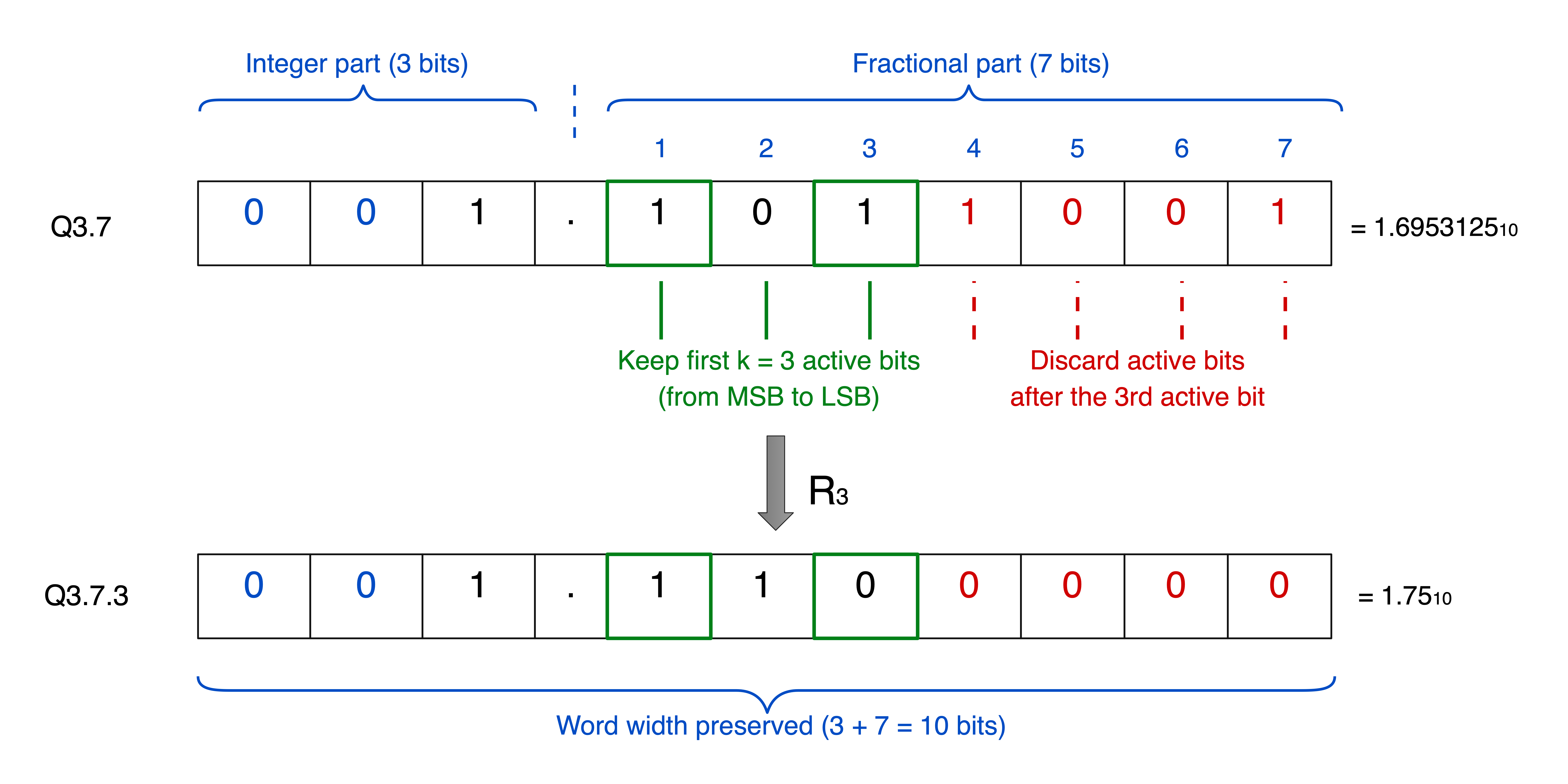}
        \label{fig:rounded_1}
    }
    \subfloat[$R_3 : \Qset{3}{7} \rightarrow \Qset{3}{7}{3}$]{%
        \includegraphics[width=0.45\linewidth]{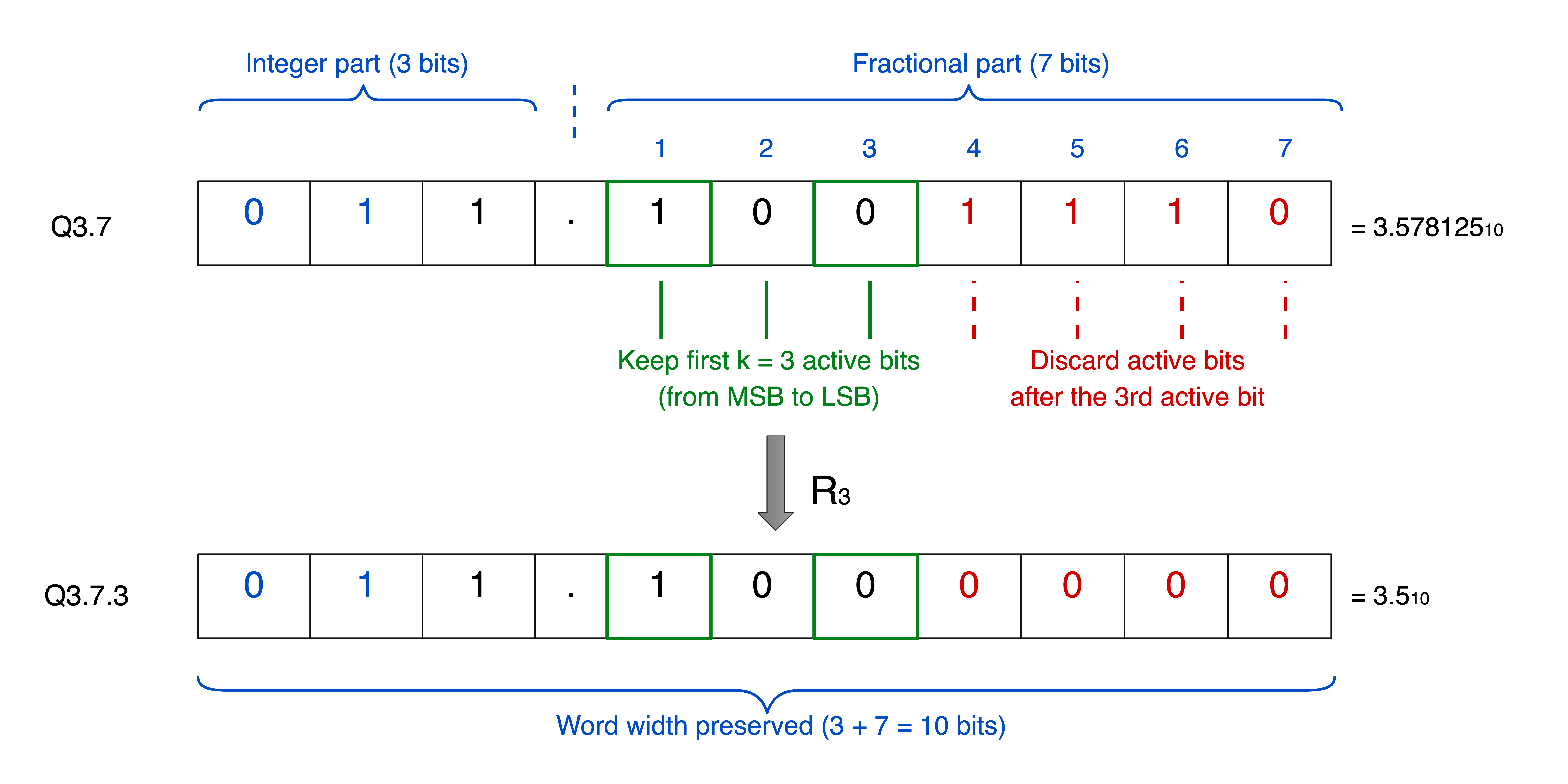}
        \label{fig:rounded_2}
    }\\
    \subfloat[$S_2 : \Qset{3}{7} \rightarrow \Qset{3}{7}{3}$]{%
        \includegraphics[width=0.45\linewidth]{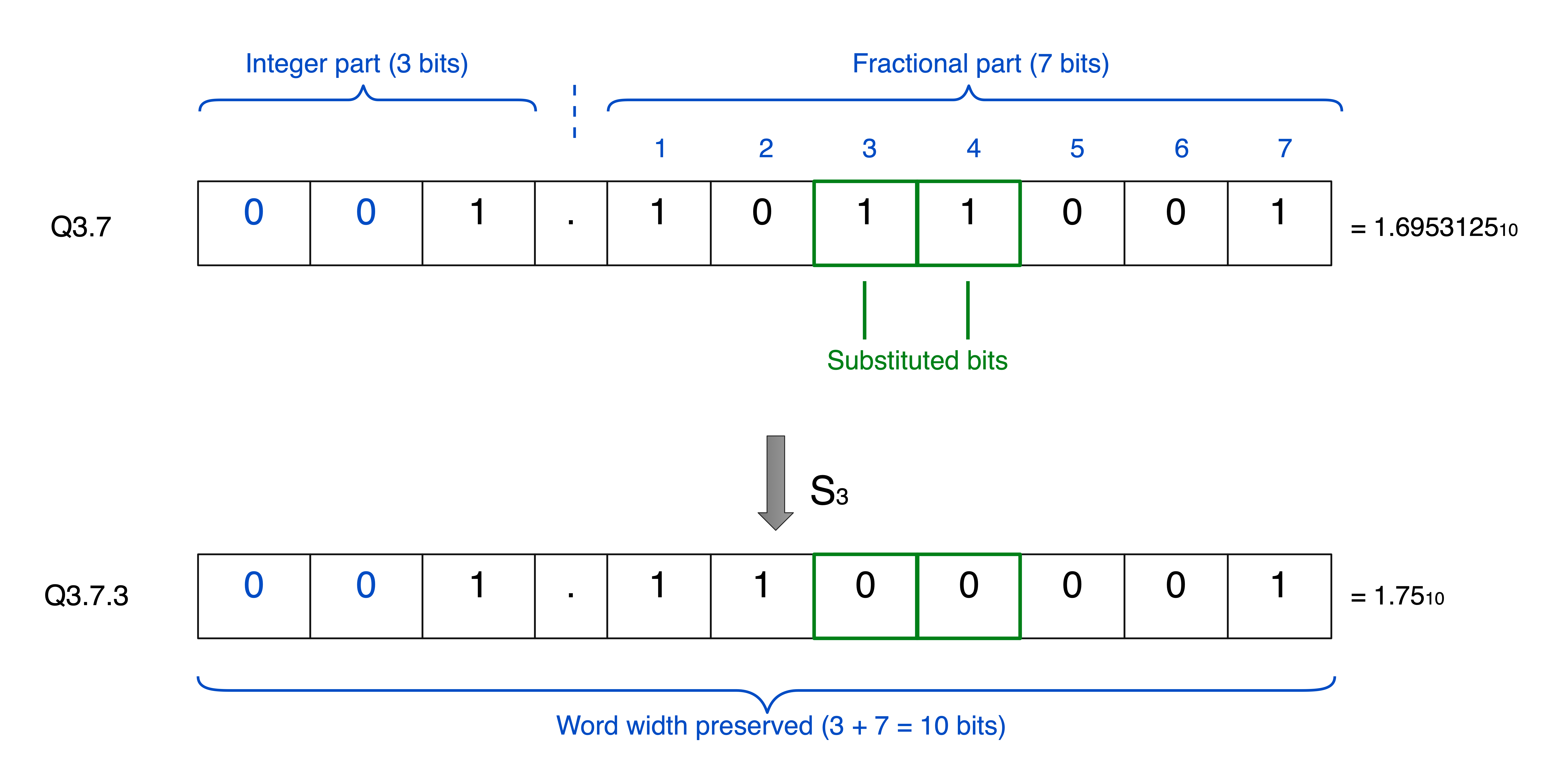}
        \label{fig:substituted_1}
    }
    \subfloat[$S_2 : \Qset{3}{7} \rightarrow \Qset{3}{7}{3}$]{%
        \includegraphics[width=0.45\linewidth]{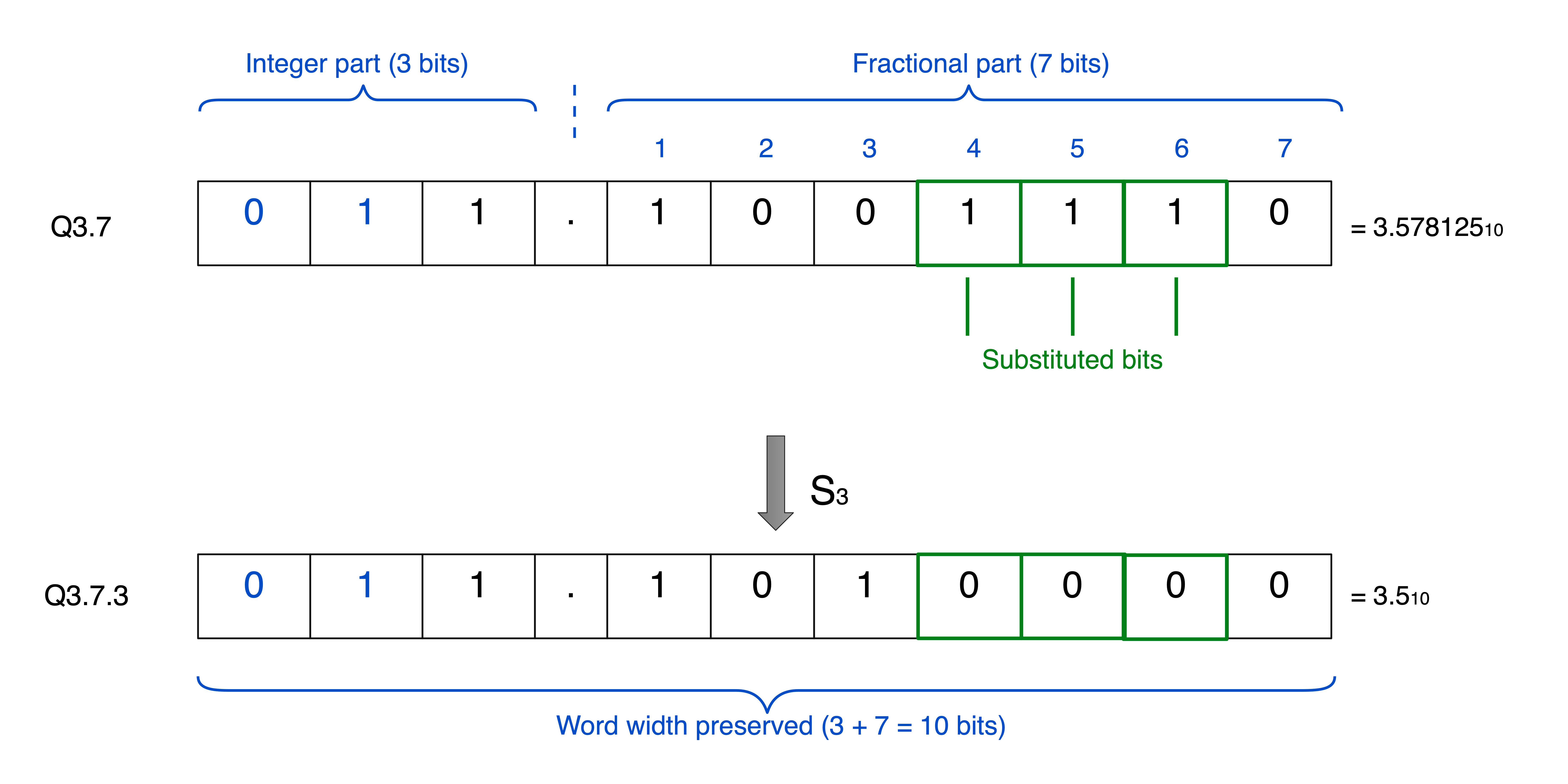}
        \label{fig:substituted_2}
    }
    \caption{Examples of transformations from \Qset{3}{7} to \Qset{3}{7}{3}. Discarded active bits are shown in red.}
    \label{fig:transformations}
\end{figure*}

\subsection{Binary Representation Design Space}

When applying aggressive active-bit reduction, the underlying binary representation significantly influences the distribution and values of the resulting quantization noise.

Two’s complement (C2) is the standard representation used in digital arithmetic due to its compact implementation of addition and multiplication operations. However, because the sign and magnitude are intrinsically coupled, negative values typically contain a larger number of active bits than positive values with comparable magnitude, particularly in the integer/sign-extension region. As a consequence, enforcing the same Hamming-weight constraint requires removing more active bits from negative values, producing larger quantization errors and a sign-dependent error distribution. These errors create a bias that accumulates during the repeated multiply-accumulate operations of the NN.

In contrast, SM separates the sign bit from the magnitude field. Therefore, positive and negative values with the same magnitude share the same active-bit pattern in the magnitude field.  Active-bit reduction only modifies the magnitude component, producing the same quantization error for both signs, yielding a  symmetric error distribution around zero. The constrained-Hamming representation consequently introduces comparable quantization errors. This symmetric behavior significantly reduces the accumulation of systematic bias during NN inference, which becomes particularly important under aggressive active-bit constraints.

Nevertheless, SM arithmetic introduces a moderate hardware overhead compared to C2, especially for addition and accumulation operations, since sign handling and magnitude comparison must be performed explicitly. Therefore, selecting the optimal representation requires trading-off the arithmetic efficiency of C2 against the improved balanced quantization-error behavior of SM under Hamming-weight constraints.

Each combination of binary representation, C2 and SM, and reduction mechanism, truncation $T_k$, rounding $R_k$ and substitution $S_l$ produces a different constrained-Hamming encoding of the original weights. Consequently, the resulting quantized values exhibit different numerical-error distributions, active-bit patterns, and arithmetic complexity characteristics. Therefore, determining the optimal constrained-Hamming configuration requires evaluating all representation-transform pairs using a criterion capable of jointly assessing numerical fidelity and hardware efficiency.

The constrained-Hamming representation already reduces the switching activity of the circuit by decreasing the number of active bits involved in the arithmetic operations, even when the same functional units are preserved. Therefore, part of the energy reduction can be obtained without introducing any specific architectural optimization, since fewer bit transitions are propagated through the existing datapath.

However, additional energy savings can be achieved by designing functional units that explicitly exploit the reduced active-bit cardinality of the operands. This is particularly relevant for hardware multipliers, which are among the most power-hungry components in NN inference accelerators. In this context, each active bit in a weight corresponds to a shifted partial product. Therefore, reducing the number of active bits directly limits the number of partial products that must be generated and accumulated.

A straightforward implementation would store the constrained-Hamming weights in their conventional binary form and let the multiplier detect the active-bit positions at run time. However, this requires additional combinational logic, including bit-scanning circuitry, priority encoders, and control logic to identify the locations of the active terms. Although functionally correct, this approach increases switching activity, hardware complexity, and energy consumption, partially offsetting the benefits introduced by the constrained-Hamming representation.

To eliminate this overhead, we propose a positional weight encoding strategy in which the locations of the active bits are precomputed offline and stored directly in memory together with the sign information. Instead of storing the complete binary representation of the weight, the encoded format stores only the positions of the active bits required by the arithmetic unit.

This transformation converts the active-bit detection problem into a direct positional lookup operation, allowing the multiplier-less arithmetic unit to generate the required shifted partial products without performing runtime bit analysis. Consequently, the proposed encoding strategy simplifies the arithmetic datapath, reduces switching activity, and improves the overall energy efficiency of the constrained-Hamming inference architecture.

\subsection{Constrained-Hamming Arithmetic Architecture} \label{sec:multiplier_architecture}

Although the proposed constrained-Hamming representation can potentially simplify several arithmetic operations within the NN datapath, its largest impact is observed in multiplication units, which dominate the computational and energy cost of NN inference accelerators. In conventional fixed-point multipliers, each active bit in the weight operand generates a partial product that must be shifted and accumulated. Consequently, reducing the active-bit cardinality directly limits the number of generated partial products, enabling substantial simplification of the arithmetic datapath.

Based on this observation, the proposed arithmetic architecture focuses on redesigning the multiplication stage to explicitly exploit the constrained-Hamming representation and the positional weight encoding introduced in the previous section.

Building upon the constrained-Hamming \Qnmk\ representation and the positional encoding strategy introduced in the previous sections, the proposed architecture replaces conventional dense multiplications with a lightweight shift-and-add arithmetic structure driven directly by the positional encoding of the active bits.

\begin{figure*} [htp]
    \centering
    \includegraphics[width = 0.8\linewidth]{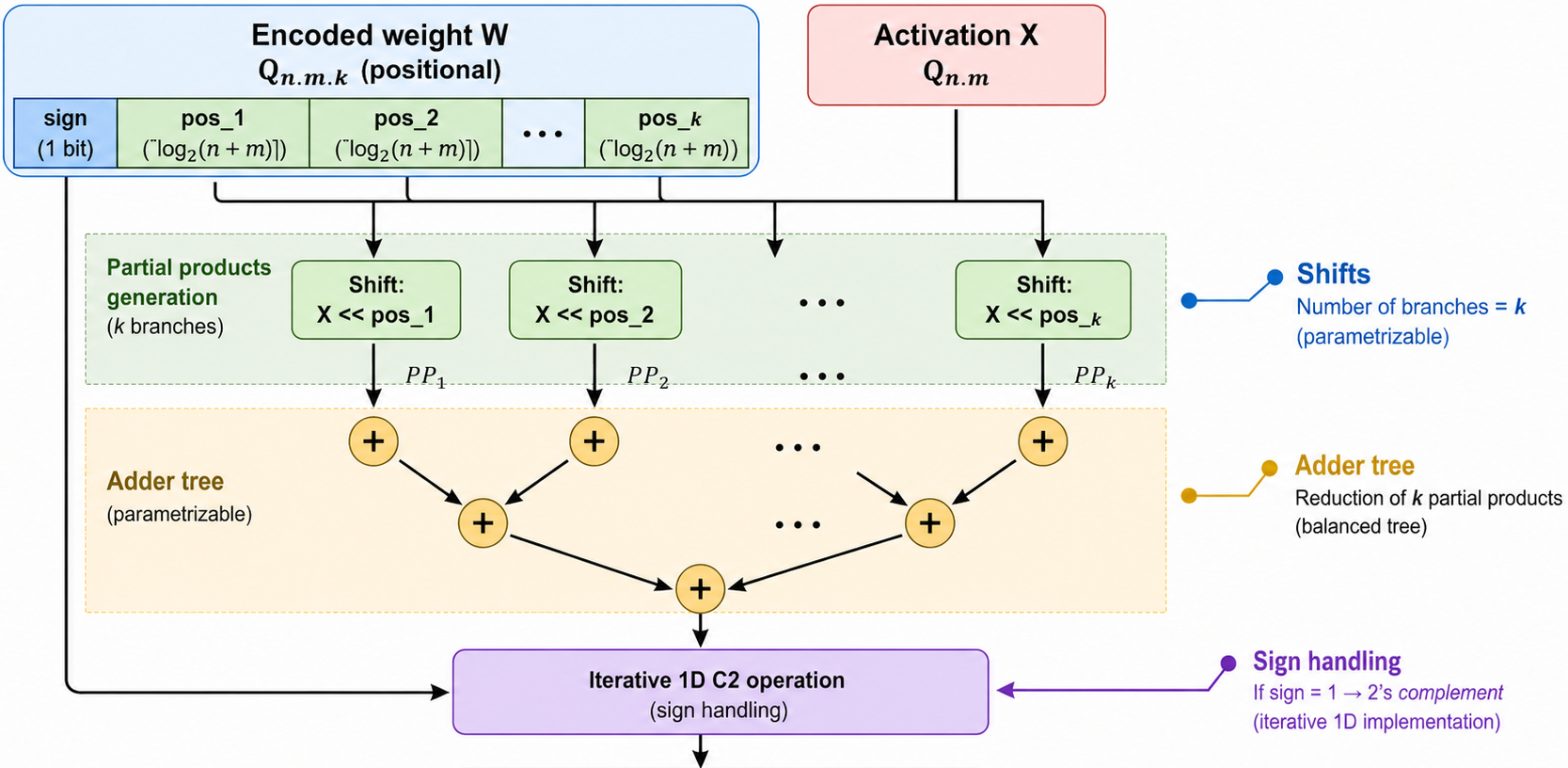}
\caption{Block-level organization of the proposed constrained-Hamming arithmetic architecture, illustrating how standard hardware multiplications are eliminated to reduce energy consumption.}
    \label{fig:flowchart}
\end{figure*}

\Cref{fig:flowchart} presents the block-level organization of the proposed constrained-Hamming arithmetic architecture. The design exploits the positional encoding introduced in the previous section to replace conventional dense multiplication with a parameterizable shift-add structure driven directly by the active-bit locations of the encoded weights.

The architecture receives two inputs: the activation operand X, represented using a conventional \Qnm fixed-point format, and the encoded weight W, represented in constrained-Hamming positional form \Qnmk. The encoded weight contains one isolated sign bit together with $k$ positional fields, where each field stores the location of one active bit within the \Qnmk\ representation. Because the sign information is handled independently from the positional fields, the internal arithmetic only operates on magnitude-driven shifts, decoupling the sign processing from the partial-product generation stage.

The first stage of the architecture corresponds to the partial-products generation network. Each positional field controls an independent shift branch that generates one shifted version of the activation operand. Consequently, the number of generated partial products is directly bounded by the active-bit constraint $k$. Since the architecture is fully parameterizable, the number of shift branches scales automatically with the selected constrained-Hamming configuration. Empty positional entries are ignored by the combinational logic, preventing unnecessary arithmetic activity and avoiding redundant switching transitions.

The generated partial products are then reduced through a parameterizable balanced adder tree. In contrast to conventional multipliers, where the number of partial products depends on the complete operand width, the proposed architecture only accumulates the partial products associated with the active-bit positions encoded in the constrained-Hamming representation. Therefore, reducing the constrained-Hamming weight directly reduces the arithmetic depth, switching activity, and logic utilization of the accumulation network. Consequently, the arithmetic complexity scales with $k$ rather than with the original operand width.

Internally, the architecture operates entirely as combinational logic without introducing additional pipeline stages, enabling direct integration into existing NN datapaths as a replacement for conventional fixed-point multipliers.

The final stage performs the sign handling operation. Once all partial products have been accumulated, the isolated sign bit controls an iterative bitwise implementation of the C2 negation. This stage applies the required sign correction without affecting the internal positional arithmetic structure, preserving the simplified shift-add datapath generated by the constrained-Hamming encoding.

Overall, the proposed architecture transforms the multiplication process into a parameterizable positional arithmetic structure whose complexity scales with the maximum number of active bits, $k$, rather than with the original operand width. This enables substantial reductions in arithmetic complexity and switching activity while preserving compatibility with conventional fixed-point NN datapaths.

\subsection{Automated RTL Design Flow} \label{sec:rtl_flow}

To evaluate the proposed arithmetic optimizations, the physical implementation is managed through a custom generation platform. The target architecture consists of a parameterizable sequence of LSTM layers followed by dense output layers. Rather than manually designing the RTL for each NN configuration, the platform automates the transition from a high-level software model to a deployable hardware description.

The process, depicted in \Cref{fig:flowchart_flow}, begins with the training of the user-defined NN architecture using standard floating-point precision to maximize accuracy. Once the model meets the required performance metrics, the quantization phase is executed, converting weights and activations to the predefined fixed-point format. Following this software-level validation, the platform proceeds to the hardware operator binding phase. At this point, the tool generates a VHDL-2008 RTL description that directly mirrors the hierarchical structure of the desired NN. This structural generation avoids generic matrix multiplication engines by instantiating specific hardware components for each operation, allowing the automated integration of our specialized multiplier-less arithmetic. Finally, the software tool automatically translates the quantized weights into the 13-bit encoding, writing them into memory initialization files to enable immediate RTL simulation and subsequent FPGA synthesis.

\begin{figure}[htp]
    \centering
    \includegraphics[width = 0.8\linewidth]{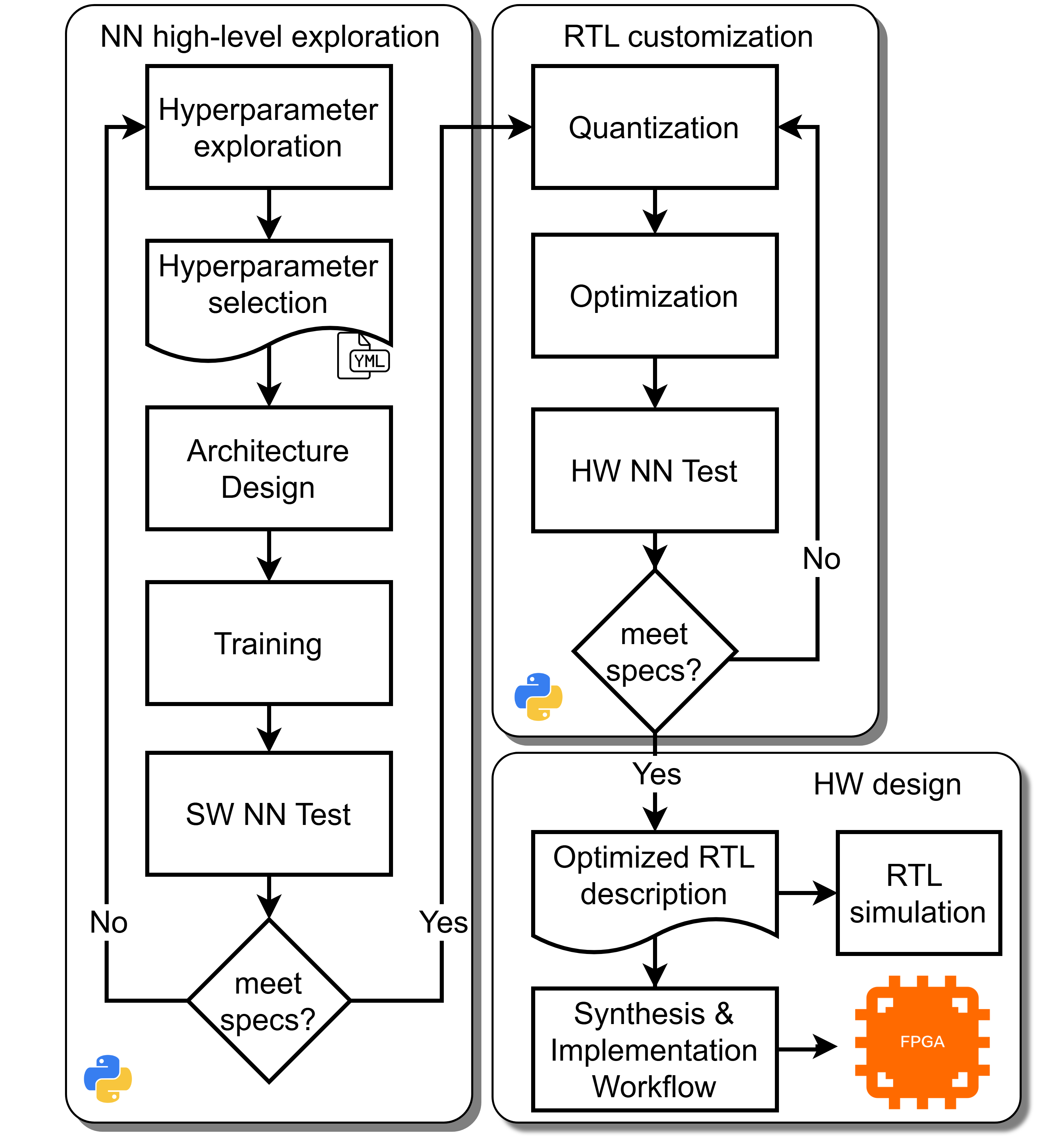}
    \caption{Complete design process, from the initial software training and quantization to the final RTL generation and weight encoding.}
    \label{fig:flowchart_flow}
\end{figure}

\section{Experimental Setup}\label{sec:results}

\subsection{Case Study and Neural Network Architecture}

Although the proposed CHQ is independent of the underlying NN architecture and applicable to a broad class of inference accelerators, a blood glucose prediction model based on LSTM was selected as the primary use case to evaluate its performance \cite{mirshekarian2017}. The specific predictive model selected for this case study consists of a single LSTM layer composed of five neurons, followed by a dense output layer. This network is designed to process an input sequence of four distinct physiological features over a temporal window of twelve timesteps. This compact structure, previously validated in \cite{tena2023}, was explicitly chosen to evaluate the hardware efficiency of the CHQ arithmetic in a highly constrained scenario.} 

This specific use case was chosen because medical monitoring wearables demand a practical balance between predictive capacity and strict computational limits. These devices operate under constant power, latency, and privacy constraints that justify deploying intelligence directly at the edge. By processing sensitive physiological data locally, edge accelerators ensure real-time forecasting without cloud connectivity, preserving patient privacy and eliminating the energy overhead of wireless data transmission. To physically evaluate this approach, the corresponding hardware implementations are automatically generated using the RTL design flow introduced in \Cref{sec:rtl_flow}.

Beyond these system-level constraints, the algorithmic nature of small scale LSTMs presents an additional challenge for quantization techniques. Unlike large convolutional networks or transformers, which exhibit significant structural redundancy and can easily absorb quantization noise, shallow LSTMs lack this inherent buffer. In these compact architectures, each individual parameter becomes critical for capturing temporal dependencies. Therefore, this specific model provides a rigorous testbed to verify that CHQ reduces energy consumption without causing a critical degradation in predictive accuracy.

To understand the computational workload of this architecture and the impact of the CHQ optimization, it is necessary to analyze the operations executed within a single LSTM cell. The internal operations of an LSTM cell at timestep $t$ are defined by \Crefrange{eq:lstm_f}{eq:lstm_h}. In these equations, $x_t$ represents the input vector, $h_{t-1}$ is the previous hidden state acting as short-term memory, and $c_t$ is the cell state vector acting as long-term memory. The flow of information is regulated by three distinct gates. The forget gate, $f_t$, identifies and discards irrelevant information from the past. The input gate, $i_t$, determines the new information to be stored in the cell state. The output gate, $o_t$, calculates the new hidden state, $h_t$, which is passed to the next iteration. These gates operate using two activation functions: the recurrent sigmoid function, $\sigma$, and the hyperbolic tangent function, $\tanh$. Mathematically, $\odot$ denotes the element-wise multiplication.

\begin{align}
    f_t &= \sigma(W_f x_t + U_f h_{t-1} + b_f) \label{eq:lstm_f} \\
    i_t &= \sigma(W_i x_t + U_i h_{t-1} + b_i) \label{eq:lstm_i} \\
    \tilde{c}_t &= \tanh(W_c x_t + U_c h_{t-1} + b_c) \label{eq:lstm_c_tilde} \\
    o_t &= \sigma(W_o x_t + U_o h_{t-1} + b_o) \label{eq:lstm_o} \\
    c_t &= f_t \odot c_{t-1} + i_t \odot \tilde{c}_t \label{eq:lstm_c} \\
    h_t &= o_t \odot \tanh(c_t) \label{eq:lstm_h}
\end{align}

Furthermore, $W_g$ and $U_g$ represent the input and recurrent weight matrices, and $b_g$ defines the bias vectors, where $g \in \{f, i, c, o\}$ designates the specific gate or cell state being computed. To quantify the computational volume, the dimensions of these matrices must be considered. Given an input vector $x_t$ with $I$ features and a hidden state $h_{t-1}$ of size $H$, each input weight matrix $W_g$ has dimensions of $H \times I$, and each recurrent weight matrix $U_g$ has dimensions of $H \times H$. From an implementation perspective, these gate operations rely on dense dot products. As demonstrated by \Crefrange{eq:lstm_f}{eq:lstm_o}, generating a single prediction requires eight dense matrix multiplications per timestep. At the hardware level, this matrix arithmetic translates to $4\cdot H \cdot (I + H)$ individual scalar multiplications for every single timestep processed by the cell. Consequently, processing an entire input sequence of length $T$ requires a total of $4TH(I + H)$ scalar multiplications to complete a single inference. Because these multiplications represent the core computational workload of the neuron, evaluating the CHQ arithmetic under this operation scheme provides a valid assessment of its energy reduction capabilities.

Applying this theoretical foundation to the selected NN architecture, we can determine its exact computational requirements. Given the established parameters of $H = 5$ neurons, $I = 4$ physiological features, and a sequence length of $T = 13$ timesteps, generating a single prediction requires $4 \times 13 \times 5 \times (4 + 5) = \num{2340}$ scalar multiplications. By completely replacing these \num{2340} standard hardware multiplications with the proposed CHQ arithmetic per prediction, the architecture aims to significantly reduce the energy consumption of the wearable device.

To ensure a highly efficient hardware baseline before applying the CHQ arithmetic, the design incorporates specific architectural optimizations.ç First, the architecture employs a sequential processing strategy where a single parameterizable neuron core is reused across different layers, prioritizing reduced resource usage over processing speed. Second, the standard exponential activation functions, $\sigma$ and $\tanh$, are replaced by linear approximations, eliminating complex transcendental calculations \cite{tena2023}. Finally, the structural organization of the neuron is mapped onto a specialized systolic array architecture. This configuration allows the intensive multiplications associated with the static $W$ and $U$ weight matrices to be executed exclusively through the proposed multiplierless arithmetic, optimizing the hardware resources without altering the functional logic of the LSTM gates.

\subsection{Dataset and Numerical Configuration}

To ensure a fair comparison across all implementations, we employed the second cohort of the OhioT1DM dataset \cite{ohio}. Missing values were completed using cubic spline interpolation.

The models were configured to predict BG levels 30 minutes ahead using a two-hour historical window sampled every 10 minutes, generating 12 samples per inference for four features: BG, carbohydrate intake, basal insulin, and insulin boluses. BG levels were scaled down by a factor of 100 to an operational range of 0.4 to 4.0, and the remaining features were normalized between 0 and 1 using Min-Max scaling.

During the initialization phase, the NN parameters were trained using an 80/20 data split with a 10-fold cross-validation approach. The training utilized the Adam optimizer with a 0.01 learning rate and MSE lossfor 100 epochs. incorporating an early stopping patience of 10 epochs. Network predictions were bounded within \SIrange[round-precision=0]{40}{400}{\BGunit} to mimic physical sensor constraints.

Finally, all parameters were quantized into a 10-bit \Qformat{3}{7} fixed-point format \cite{tena2021} for hardware deployment. This configuration balances clinical accuracy and hardware logic overhead.

\section{Representation-Transformation Analysis}\label{sec:representation_analysis}

To explore the design space defined by the combinations of binary representations (C2 and SM), transformation mechanisms ($T_k$, $R_k$, and $S_\ell$), and active-bit constraint $k$, two complementary metrics are employed: numerical fidelity, measured through the Signal-to-Noise Ratio (SNR), and arithmetic complexity, estimated through the Hamming weight of the transformed representation.

Numerical fidelity is evaluated through the SNR introduced by the constrained-Hamming representation. Let $X$ denote a set of original fixed-point values in \Snm and $\widehat{X}$ their transformed representations in \Snmk. The quantization noise is defined as $X-\widehat{X}$, and the resulting SNR is computed as:

\begin{equation}
\text{SNR}=10 \cdot \log_{10}\left(\frac{\sum_i X_i^2}{\sum_i (X_i-\widehat{X}_i)^2}\right) \label{eq:snr}
\end{equation}


Higher SNR values (in dB) indicate a closer numerical approximation to the original representation and therefore lower quantization-induced distortion.

Hardware efficiency is estimated through the aggregated active-bit cardinality of the transformed weight set, $H_{\text{tot}}$, i.e., the total constrained-Hamming weight obtained by summing the Hamming weight across all transformed weights. 

\begin{equation}
H_{\text{tot}} = \sum_{w \in \text{weights}} H(w) \label{eq:hw_eff}
\end{equation}

We propose a multiplier-less arithmetic where each active bit generates a shifted partial product that must be accumulated during multiplication. Consequently, reducing the number of active bits directly reduces the number of shift-and-add operations, simplifying the arithmetic hardware and decreasing switching activity, logic utilization, and energy consumption.

Therefore, the optimal constrained-Hamming configuration is determined through a Pareto exploration that jointly evaluates numerical accuracy and arithmetic complexity across all evaluated representation-transform combinations. We conduct this analysis on the input and recurrent weights, U and W matrices, respectively.

Different values of $k$ define different constrained-Hamming design spaces and for every value of $k$, there exist solutions on the Pareto front for multiple combinations of transformations. Lower values of $k$ lead to more aggressive arithmetic simplification but often introduce excessive quantization noise, whereas higher values reduce quantization noise at the expense of weaker hardware simplification. Therefore, the selection of $k$ is not driven by Pareto optimality, since solutions corresponding to all values of $k$ appear on the Pareto front. Instead, $k$ should be regarded as a design parameter that controls the trade-off between hardware simplification and quantization accuracy. In this work, $k=3$ was selected as the representative design point because it provides a practical compromise between both objectives: it significantly reduces the active-bit cardinality while avoiding the quantization-noise increase observed for more restrictive constraints.

\begin{figure}[htp]
    \centering
    \subfloat[Input weights (U matrix) ]{%
        \includegraphics[width=0.9\linewidth]{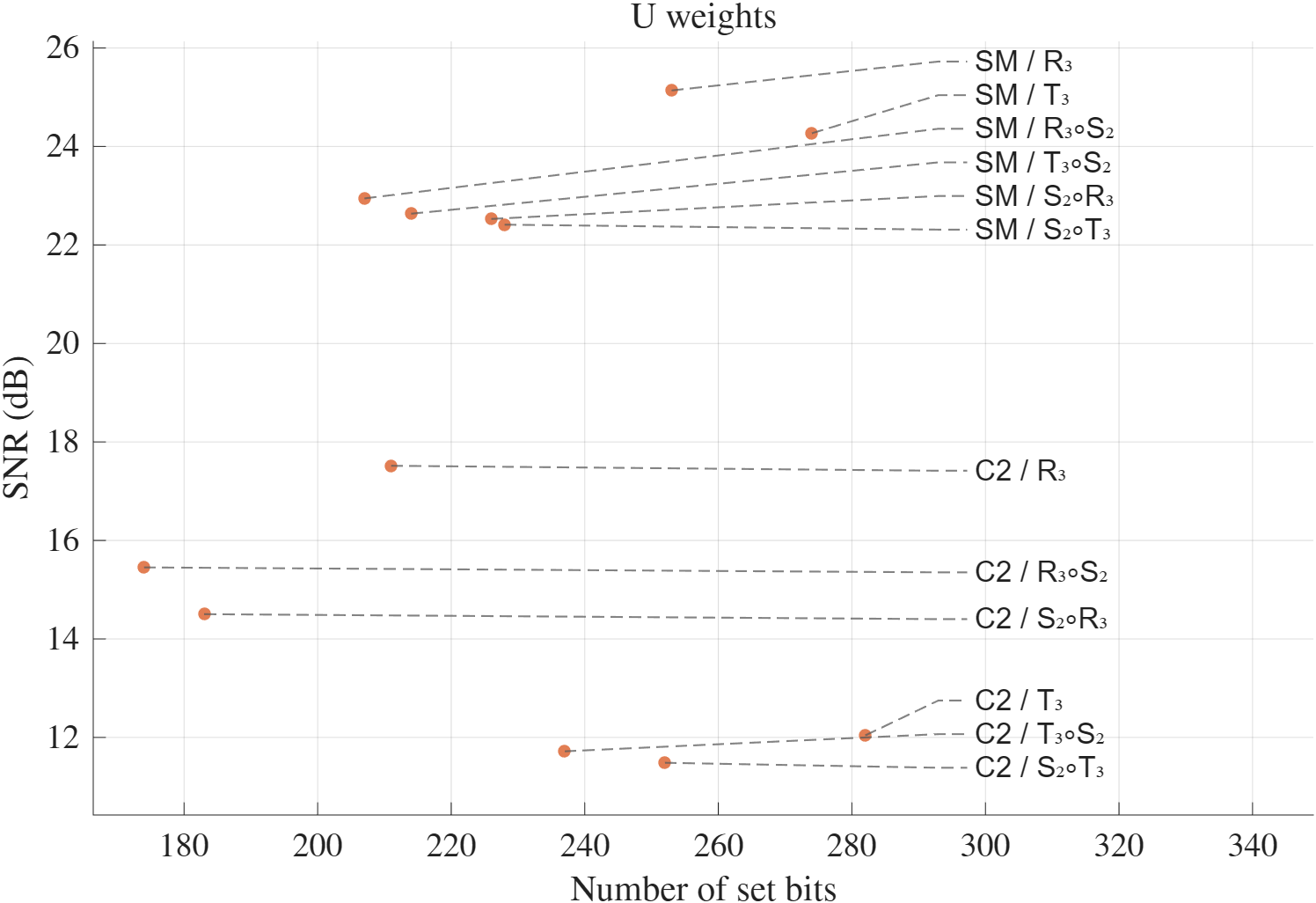}
        \label{fig:pareto_U}
    } \\
    \subfloat[Recurrent weights (W matrix)]{%
        \includegraphics[width=0.9\linewidth]{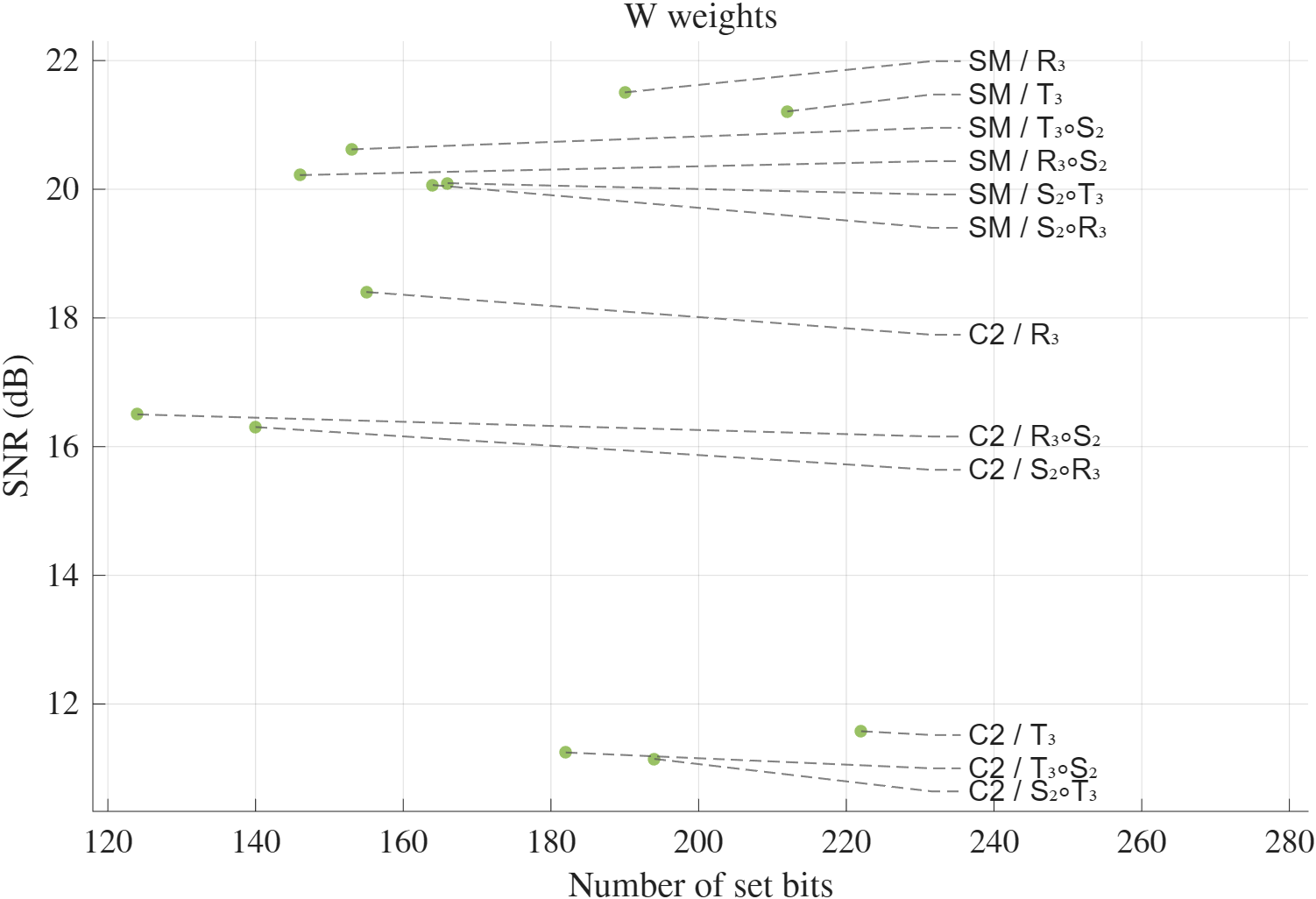}
        \label{fig:pareto_W}
    }
    \caption{Pareto analysis of the representation-transformation design space for the input and recurrent weight matrices. The horizontal axis is the aggregate constrained-Hamming weight, $H_\text{tot}$, of the transformed weights (shown as Number of Set Bits), whereas the vertical axis is the resulting SNR. Labels identify the evaluated representation-transformation combinations.}
    \label{fig:pareto_combined}
\end{figure}

 \Cref{fig:pareto_combined} illustrates the constrained-Hamming design space for the input and recurrent weight matrices for $k=3$. Numerical fidelity is measured by the resulting SNR, whereas hardware efficiency is estimated by the aggregate constrained-Hamming weight $H_{\text{tot}}$. Each point corresponds to a configuration obtained by applying a transformation $\mathcal{T} \in \{S,R_3,T_3,R_3\! \circ\! S_2,T_3 \! \circ \! S_2,S_2 \! \circ \! R_3,S_2\!\circ\!T_3\}$ under either SM or C2 representations. As illustrated in \Crefrange{fig:pareto_U}{fig:pareto_W}, the configurations evaluated in SM format, consistently populate the upper region of the Pareto space. They offer a higher SNR compared to their C2 counterparts, confirming that the symmetric error distribution of SM preserves numerical fidelity more effectively than C2, compensating for its minor logic overhead. The figures also show the impact of the $S_2$ transformation on the resulting representation. It reduces the number of active bits although this reduction is accompanied by a decrease in SNR. As a result, the corresponding configurations move towards the lower-left region of the Pareto space.

The Pareto fronts identify several non-dominated representation-transformation combinations. Among them, the configuration obtained using the SM representation and the $R_3$ operator achieves the highest numerical fidelity while still reducing the total number of active bits by approximately \SI[round-precision=0]{56}{\percent}. With a mean SNR of \SI{24.532}{\dB}, this configuration was selected for the representation of NN weights for the remainder of this work.

\section{Arithmetic Hardware Analysis} \label{sec:multiplier_implementation}

To assess the hardware implications of the proposed constrained-Hamming arithmetic architecture, the design described in \Cref{sec:multiplier_architecture} has been implemented in VHDL-2008 and compared against a conventional 10-bit \Qformat{3}{7} fixed-point multiplier automatically inferred by Vivado for the target FPGA. The evaluated constrained-Hamming configuration employs a \Qformat{3}{7}[3] representation generated through the $R_3$ transformation and SM encoding selected in the analysis presented in \Cref{sec:representation_analysis}. \Cref{tab:multiplier} reports the resulting latency, resource utilization, power consumption, energy per operation, and energy-delay product (EDP) after synthesis, place-and-route, and switching-activity-based power estimation.

Both designs were described using VHDL-2008, synthesized and implemented using Vivado 2021.1, and simulated using QuestaSim 2020.4. The synthesized designs target the Xilinx Virtex-7 XC7VX485T-2FFG1761C FPGA, available on the VC707 Evaluation Board, and built in TSMC's \Tec{28} process technology (28HPL). For energy consumption monitoring, we rely on the power analysis reports generated by Vivado. To ensure an accurate dynamic power assessment, the implementation design is simulated using realistic physiological input data to generate a Switching Activity Interchange Format file. This file captures the exact toggle rates of the internal signals, allowing the Vivado power analyzer to compute the dynamic and static power dissipation under real operating conditions. However, it is important to note that the Xilinx Virtex-7 FPGA is not our ultimate target. Although the Virtex-7 device is used as the evaluation platform, the objective of this analysis is not to target this specific FPGA family, but to compare the relative hardware cost and energy characteristics of the proposed arithmetic architecture against a conventional multiplier implementation under identical synthesis conditions.

In the Virtex-7 architecture, standard multiplications are typically mapped onto dedicated DSP48E1 blocks, which are highly optimized arithmetic resources specifically designed for multiplication and accumulation operations. Consequently, this comparison introduces an advantage in favor of the conventional implementation, since the reference multiplier benefits from dedicated silicon resources whereas the proposed architecture is implemented using general-purpose programmable logic. Therefore, any improvement achieved by the constrained-Hamming architecture should be interpreted as a conservative estimate of its potential benefits.

Unless otherwise specified, the hardware evaluation methodology described above is used consistently throughout the remainder of this work for all synthesized arithmetic units and NN implementations.

\begin{table}[htp] 
  \centering
  \caption{Hardware implementation results for the proposed constrained-Hamming arithmetic architecture and a conventional 10-bit fixed-point multiplier.}
    \begin{tabular}{l
                    S[table-format=3.2]
                    S[table-format=3.2]}
    \toprule
                                   & \textbf{Baseline} & \textbf{CHQ} \\  
    \midrule
    Latency (\si{\nano\second})  &  3.415      & 4.238   \\
    \midrule
    Slice LUTs                     & 102       & 154  \\
    FF                             & 0         & 0  \\
    Slices                         & 32        & 46  \\
    \midrule
    Power (\si{\milli\watt})     & 0.18        & 0.12  \\
    Energy/op (\si{\pico\joule})      & 2.2477  & 1.4985  \\ 
    EDP ($10^{-21}$\si{\joule\second})     &  7.6758 & 6.35064   \\
    \bottomrule
    \end{tabular}
  \label{tab:multiplier}
\end{table}

Although the proposed architecture requires additional programmable logic resources to implement the positional arithmetic datapath, the reduction in switching activity resulting from the constrained-Hamming representation compensates for this overhead, leading to a lower dynamic power consumption and improved energy efficiency. Compared with the DSP-based baseline, the proposed architecture reduces power consumption by approximately \SI{33}{\percent}, energy per operation by \SI{33}{\percent}, and EDP by \SI{17}{\percent}, at the expense of a \SI{50.98}{\percent} increase in logic utilization and a \SI{24}{\percent} increase in latency. The increased latency is expected, since the baseline implementation benefits from highly optimized DSP48E1 blocks, whereas the proposed architecture is realized using programmable logic resources.

\section{Neural Network Analysis} \label{sec:system_level}

To systematically analyze the proposed CHQ and its associated arithmetic architecture, the experimental design is structured to assess three primary metrics: predictive precision, energy consumption, and system-level scalability. First, the analysis focuses on quantifying the precision loss introduced by the arithmetic optimizations, addressing the primary challenge of the constrained-Hamming arithmetic architecture: maintaining a level of predictive performance that remains clinically useful despite the reduction in arithmetic complexity. Second, we analyze the energy consumption of the physical implementation. Finally, a comparative analysis is presented to benchmark CHQ against both our first hardware implementation proposal \cite{tena2023} and a consumer smartphone, which serves as an everyday edge device capable of hosting AI solutions such as the BG predictor. This comparison is conducted across LSTM models of incremental size to validate the scalability and efficiency of the architecture across different NN depths.

\subsection{Experimental Methodology}

For the accuracy analysis, a specific NN architecture established in previous literature \cite{mirshekarian2017}, consisting of a single LSTM layer with five neurons, was analyzed employing a 10-fold cross-validation strategy to ensure robust results. To isolate the exact error introduced by CHQ, the predictive performance of the newly proposed architecture was directly compared against the previously optimized hardware implementation detailed in \cite{tena2023}. Both implementations utilized the established \Qformat{3}{7} fixed-point format. To quantify the precision loss, the Root Mean Square Error (RMSE) was selected as the primary mathematical metric. Additionally, the Parkes Error Grid (PEG) was utilized to analyze the clinical validity and safety of the predictions. The PEG categorizes errors into five risk zones based on treatment consequences, where zones A and B represent clinically benign predictions, zone C indicates an over correction and Zones E and D indicate increasingly severe medical errors \cite{parkes}.

For the energy consumption and scalability analysis, a different set of models was generated to assess the hardware efficiency. This experimental design starts with the BG model as smallest NN, then 16 LSTM neurons are maintained per layer, alongside a single dense neuron in the output layer. The NN depth was systematically adjusted, starting from a single layer and incrementing by 25 layers up to \num{150}, followed by extended depths of 200, 250, and 500 layers. The entire workflow was executed for these ten distinct structural configurations to properly initialize the NN weights, avoiding zero multiplications that would artificially lower energy measurements and misrepresent real-world operation.

To provide a comprehensive analysis for this second analysis, the study contrasts three distinct implementations. First, the newly optimized constrained-Hamming arithmetic architecture is compared against our baseline hardware implementation on an FPGA, a comparison that isolates and quantifies the energy savings achieved directly by the proposed arithmetic modifications. 
The low energy LSTM models are developed using the platform for architecture exploration, training, analysis, and RTL generation, which is synthesized, implemented, and simulated using the methodology exposed in \Cref{sec:multiplier_architecture}.

Second, both hardware models are analyzed against an equivalent software implementation deployed on a AI-ready smartphone. Because wearable monitoring sensors are commonly integrated with mobile applications, and modern smartphones offer substantial computational capabilities, they serve as a standard and realistic baseline for deploying everyday AI models at the edge. To establish the smartphone baseline, the trained weights were extracted and the corresponding models were manually constructed in Swift. These software models were subsequently integrated into an iOS application, enabling independent testing for each NN configuration. The smartphone is an Apple\textsuperscript{\textregistered} iPhone 13 Pro\textsuperscript{TM}, operating under iOS 15.4.1, featuring an A15 Bionic chipset manufactured using TSMC's \Tec{5} (N5P) process technology. The chipset runs at a clock speed of \SI{3.23}{\giga\hertz} and is powered by a \SI[round-precision=0]{3095}{\milli\ampere\hour} battery with a nominal voltage of \SI{3.91}{\volt}. The A15 Bionic chipset includes specialized ML cores known as Apple's Neural Engines. These cores are designed to accelerate AI operations, reducing processing time while maintaining accuracy. To facilitate seamless integration, we ported the model from Python to the Swift programming language using the CoreML library. The app itself was developed using XCode, a suite of tools that encompasses user interface design, coding, testing, and debugging. For energy consumption monitoring, we rely on the iOS built-in battery usage app, which tracks the energy usage of the iOS application running the NN.

For the three analyzed implementations, specifically the baseline hardware model, the proposed constrained-Hamming arithmetic architecture, and the software deployment on the consumer smartphone, energy consumption per inference was computed using distinct methodologies based on the target platform. To accurately reflect our measurement procedures, we formulated specific equations, \Cref{eq:energyFPGA} and \Cref{eq:energyiOS}, for each hardware domain.

\begin{align}
    E^{\text{FPGA}}_{\text{inf}} = \frac{1}{f}*n_{\text{cycles}}*(W_{\text{dyn}}+W_{\text{static}})
    \label{eq:energyFPGA} \\
    E^{\text{iOS}}_{\text{inf}} = B_{T}*B_{\text{frac}}*\frac{t_{\text{inf}}}{t_{\text{exp}}}
    \label{eq:energyiOS}
\end{align}

To determine the energy per inference consumed by the two hardware models implemented in the FPGA, we established \Cref{eq:energyFPGA}. This equation was formulated specifically to utilize the data provided by the synthesis and simulation tools. In this equation, $E^{\text{FPGA}}_{\text{inf}}$ represents the energy consumed per inference, $f$ denotes the clock frequency, which is fixed at \SI[round-precision=0]{10}{\mega\hertz}, $n_{\text{cycles}}$ corresponds to the number of cycles required for a single inference, $W_{\text{dyn}}$ represents the dynamic power consumed by the circuit, and $W_{\text{static}}$ is the total static power necessary to keep the entire FPGA operational.

To analyze the energy consumption of the smartphone baseline, we developed an iOS application embedded with the corresponding LSTM models. This application features a button that triggers inferences in batches of \num[round-precision=0]{5000} entries, to deplete at least \SI[round-precision=0]{10}{\percent} of a fully charged battery, along with the average time required for a single inference. As a direct result of this testing methodology, we derived \Cref{eq:energyiOS}. In this equation, $E^{\text{iOS}}_{\text{inf}}$ represents the energy consumed per inference, $B_{T}$ denotes the total battery capacity, which is \SI{12.11}{\watt\hour}, $B_{\text{frac}}$ corresponds to the percentage of battery consumed, $t_{\text{exp}}$ signifies the duration of each experiment, i.e., the time required to process all the inferences made, and $t_{\text{inf}}$ represents the average inference time.

\subsection{Accuracy Analysis}

The initial analysis, on the testing partition, using the RMSE revealed that the standard hardware implementation, utilizing the full range of the \Qformat{3}{7} format, achieved a value of \SI{25.26}{\BGunit}. In comparison, the constrained-Hamming arithmetic architecture yielded an RMSE of \SI{29.45}{\BGunit}. This numerical increase of \SI{4.19}{\BGunit} is the result of the quantization noise and the loss of resolution. Therefore, to properly assess whether this statistical increase actually impacts medical decisions, it is necessary to perform a clinical analysis using the PEG analysis.

\begin{figure}[htp]
    \centering
    \begin{tabular}{ c }
        \subfloat[Baseline Hardware LSTM]{ %
            \includegraphics[width=0.8\linewidth]{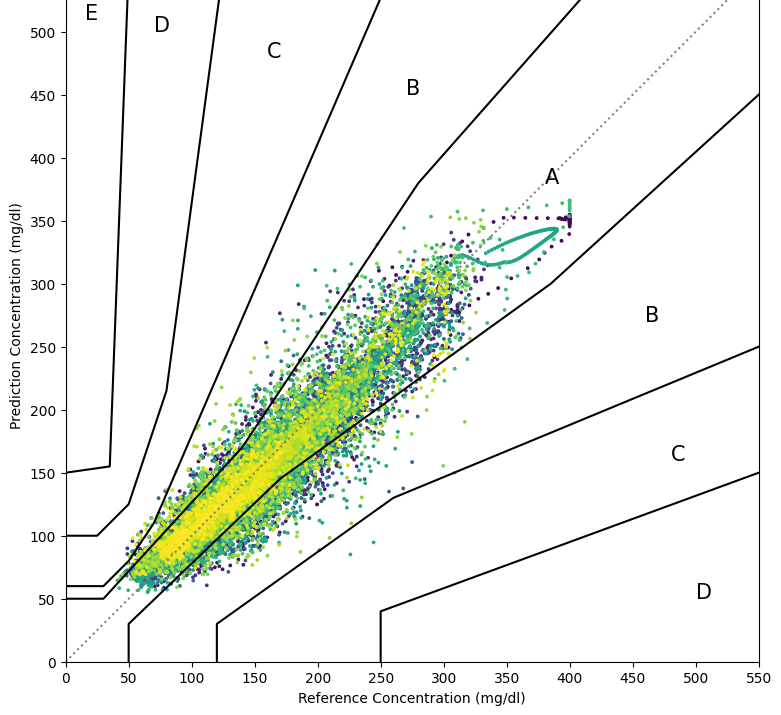}
} \\
        \subfloat[CHQ-based LSTM]{ %
            \includegraphics[width=0.8\linewidth]{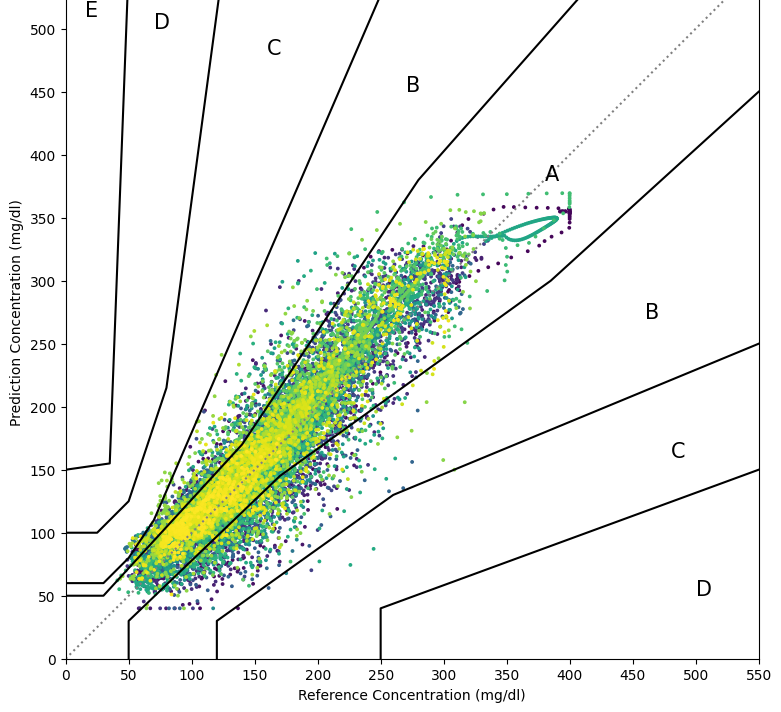}
}
    \end{tabular}
    \caption{PEG analysis for the baseline hardware LSTM (a) and the proposed constrained-Hamming arithmetic architecture (b). Both models demonstrate comparable clinical validity, maintaining the majority of predictions within the clinically acceptable zones.}
    \label{fig:peg_comparison}
\end{figure}

Following this approach, \Cref{fig:peg_comparison} shows that the distribution of the predicted values against the ground truth is similar for both models. The baseline hardware model placed \SI{98.21}{\percent} of its predictions within the clinically benign Zones A and B, which represent points where treatment decisions based on the model would lead to little or no negative clinical consequences. The constrained-Hamming arithmetic architecture maintained a similar performance, with \SI{98.07}{\percent} of its predictions falling within the same zones, although \SI{4}{\percent} of the points have shifted from zone A to zone B.

This difference of only \SI{0.14}{\percent} in the clinically benign regions, zones A and B, demonstrates that the CHQ preserves the predictive features of the NN. The trade-off between mathematical precision and hardware efficiency indicates that the reduction in computational overhead is achieved without compromising the clinical utility of the medical wearable device.

\subsection{Energy and Scalability Analysis}

The comparative analysis of energy consumption per inference begins with the specific BG model configuration, which consists of 1 layer with five LSTM neurons and an output layer with one dense neuron. When operating with this configuration, the proposed CHQ architecture consumes \SI{3956}{\nano\joule}, compared to the \SI{4950}{\nano\joule} consumed by the original baseline FPGA implementation. This achieves a \SI{25}{\percent} energy savings within the FPGA environment.

\begin{figure}[htp]
    \centering
    \subfloat[Energy comparison between the baseline and the CHQ FPGA  incremental implementations in logarithmic scale.]{%
        \includegraphics[width=0.9\linewidth]{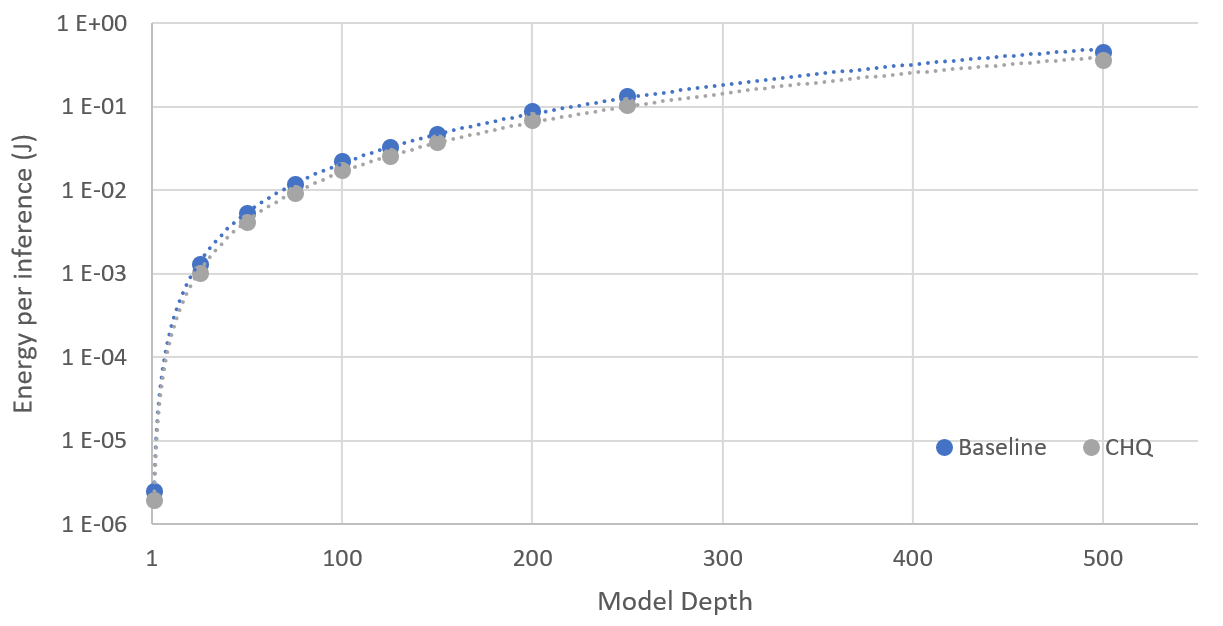}
        \label{fig:energy_hws}
    } \\
    \subfloat[Comparison of relative energy savings between the baseline and the CHQ hardware implementations.]{%
        \includegraphics[width=0.9\linewidth]{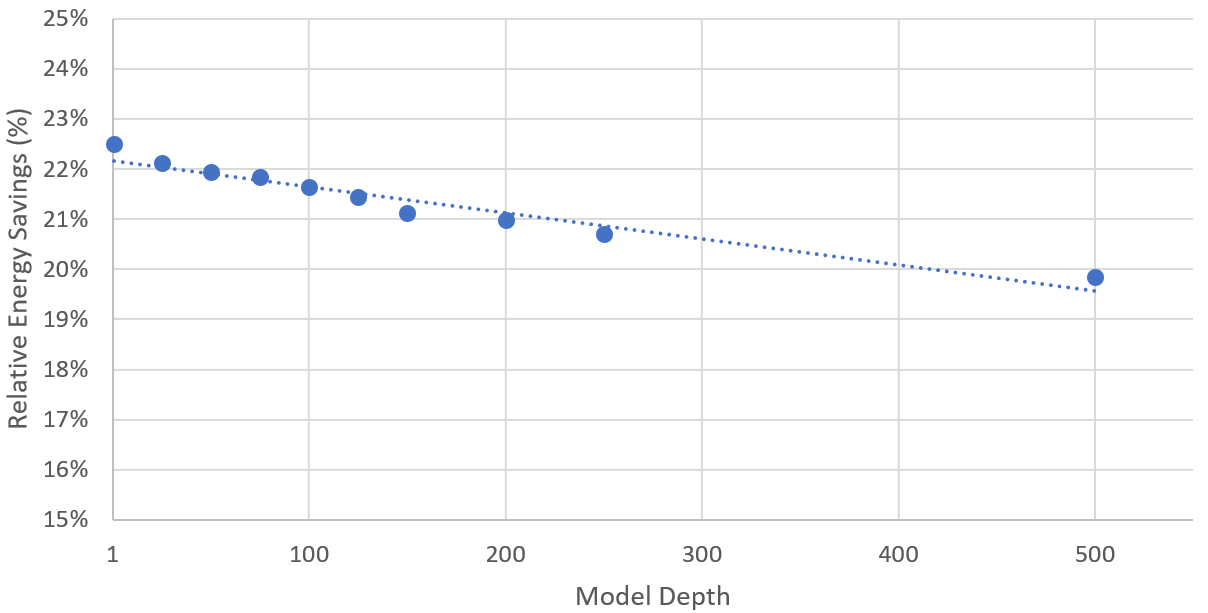}
        \label{fig:energy_rel}
    }
    \caption{Energy consumption per inference as a function of the number of LSTM layers. The plots illustrate the absolute energy consumption in logarithmic scale (a) and the relative energy savings introduced by the CHQ architecture (b).}
    \label{fig:energy_double}
\end{figure}

Following this initial assessment, the analysis expands to the incremental models to evaluate scalability within the hardware domain. This comparison is detailed across two plots in \Cref{fig:energy_double}. The first plot, \Cref{fig:energy_hws}, compares the baseline and the CHQ FPGA implementations to evaluate the absolute energy reduction across the varying NN depths.  In the most complex configuration, with 500 layers, comprising a total of \num{8000} LSTM neurons, the baseline hardware model has a consumption of \SI{0.44436}{\joule} and the CHQ architecture reduces this to \SI{0.3563}{\joule}.

The second plot, \Cref{fig:energy_rel}, illustrates the analysis of energy trends. Across the evaluated configurations, the proposed architecture provides a mean energy reduction of \SI{21.39}{\percent}, with a standard deviation of \SI{0.78}{\percent}. However, the chart reveals that the relative savings achieved by the CHQ architecture diminish as the network scales. This dilution of efficiency is driven by the components that remain unoptimized. First, when increasing the number of neurons per layer from five to 16, the  multiplications associated with the hidden state weight matrices grow quadratically. This reduces the relative proportion of total energy dedicated to the optimized weight multiplications, causing the initial savings to drop from \SI{25}{\percent} to \SI{22.48}{\percent} within a single layer. In contrast, when scaling the network depth up to 500 layers while maintaining 16 neurons per layer, the number of operations grows linearly. This linear scaling preserves the relative energy ratio between the multipliers and the rest of the circuit. Therefore, the subsequent relative efficiency reduction of \SI{2.66}{\percent} between the model with one layer and the model with 500 layers is not due to a shift in algorithmic proportions, but rather to physical implementation constraints. Since the CHQ unit occupies a larger physical area than a standard multiplier, expanding the architecture introduces longer routing paths on the FPGA. These extended paths, combined with the continuous memory access and static power required to sustain a larger circuit, gradually dilute the overall energy savings.

\begin{figure}[htp]
    \centering
    \includegraphics[width=0.9\linewidth]{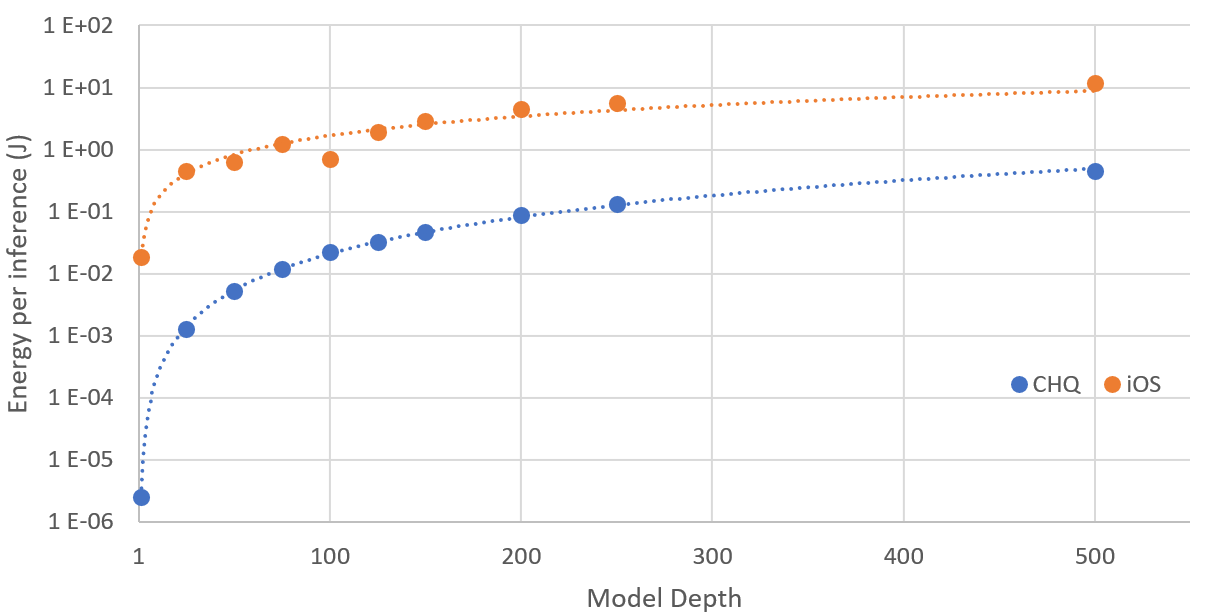}
    \caption{Energy scaling comparison between the iOS implementation and the CHQ FPGA implementation.}
    \label{fig:energy_iOS}
\end{figure}

To further contextualize these hardware efficiency gains, \Cref{fig:energy_iOS} compares the energy consumption of the proposed CHQ architecture directly against the iOS software baseline across the varying NN depths. For the initial BG model configuration, the iOS device consumes \SI{34763}{\nano\joule} per inference, whereas the CHQ architecture requires only \SI{3956}{\nano\joule}, resulting in an efficiency factor of approximately $\times$\num[round-precision=1]{8.8}. As the computational complexity increases, this efficiency gap widens. In the most complex configuration with \num{500} layers, the iOS model consumes \SI{11.686}{\joule}, while the CHQ implementation requires \SI{0.3563}{\joule}, consuming approximately $\times$\num[round-precision=1]{32.8} less energy. This trend demonstrates that the CHQ methodology provides scalable energy savings that become increasingly evident in deeper architectures.

Regarding computational performance, both hardware implementations require identical execution times for a inference across all evaluated configurations. As previously detailed in \Cref{tab:multiplier}, the CHQ unit introduces a latency increase at the arithmetic level, rising from \SI{3.415}{\nano\second} to \SI{4.238}{\nano\second}. However, this local delay does not affect the system-level inference time, because the paths containing the CHQ components do not become the critical path of the architecture. Consequently, the system achieves the reported energy savings without incurring any penalty in processing speed.

\section{Conclusions}\label{sec:conclusion}

In this study, we present the CHQ approach accompanied by a multiplier-less arithmetic architecture designed to minimize energy consumption in energy-constrained edge devices. While this framework is inherently adaptable to diverse architectures, it has a notable impact on algorithms dominated by multiply-accumulate operations, such as the LSTM focused on blood glucose forecasting utilized as our case study.
To achieve this efficiency, CHQ limits the number of non-zero bits within the fixed-point representation of the NN weights to a predefined maximum. This constrained active-bit cardinality enables a positional encoding strategy that replaces standard hardware multipliers with an arithmetic structure based on shifts and additions, thereby reducing switching activity and dynamic power dissipation. To validate the proposed framework, an automated design flow was employed to execute the entire pipeline, from software training and quantization to weight encoding and final RTL generation. This approach facilitates the exploration of different quantization configurations and enables a comprehensive multi-level evaluation. By assessing representation fidelity, arithmetic hardware utilization, and system-level energy consumption, the results confirm that the CHQ architecture meets the required performance indicators for edge deployment, providing a practical balance between predictive accuracy and energy reduction, as detailed in the subsequent paragraphs.

First, a comprehensive representation and transformation analysis was conducted to systematically explore the quantization design space and establish the core constrained-Hamming configuration. This evaluation demonstrated that SM encoding, when paired with a rounding reduction mechanism, effectively isolated quantization noise, providing the optimal balance between preserving numerical fidelity and achieving arithmetic simplification. By actively limiting the words to a maximum of three non-zero terms ($k=3$), CHQ drastically reduced the total number of active bits by \SI{56}{\percent}. Despite this aggressive mathematical compression, the representation successfully maintained a robust SNR of \SI{24.532}{\dB}, proving that complex parameter distributions can be heavily constrained without loosing information. 

Building directly upon this optimized CHQ representation, the arithmetic hardware evaluation confirmed the architectural benefits of CHQ. By translating the constrained active-bit cardinality into a positional encoding, the proposed multiplier-less unit successfully replaced conventional DSP-based fixed-point multipliers. This hardware transformation reduced power consumption from \SI{0.18}{\milli\watt} to \SI{0.12}{\milli\watt} and achieved a substantial decrease in energy per operation, dropping from \SI{2.2477}{\pico\joule} to \SI{1.4985}{\pico\joule}. 

Finally, a system-level accuracy evaluation was conducted using BG prediction as a test environment. Here, PEG analysis confirmed that \SI{99.70}{\percent} of the CHQ predictions remained within clinically benign zones, virtually matching the non-optimized implementation's \SI{99.84}{\percent}, confirming the preservation of enough precision for sensitive edge applications while delivering energy reductions.

To assess scalability, incrementally sized LSTM models were evaluated across a consumer smartphone (Apple\textsuperscript{\textregistered} iPhone 13 Pro\textsuperscript{TM}) and a Xilinx Virtex-7 FPGA hosting both the baseline and constrained-Hamming arithmetic architectures. The evaluation demonstrated that efficiency differences between software and hardware implementations widen as model complexity increases. In 500-layer configurations, the constrained-Hamming arithmetic architecture consumed $\times$\num[round-precision=1]{32.8} less energy than the iOS baseline, improving upon the $\times$\num[round-precision=1]{26.3} reduction of our first hardware architecture \cite{tena2023}. Furthermore, the hardware models consistently reduced computational latency, completing inferences nearly five times faster than the software baseline in the most complex scenarios.

This comparison provides a concrete representation of the efficiency gains achieved through the CHQ optimization strategy. Specifically, as a consequence of the CHQ constraints, the resulting multiplier-less arithmetic yields scalable energy savings that become more prominent as the computational complexity of the NN increases. This trend confirms that the constrained-Hamming arithmetic architecture is a suitable alternative for energy-constrained edge devices, which operate under strict battery capacity and size limitations.

Ultimately, CHQ establishes a representation framework that enables the deployment of complex time-series forecasting models with a minimal energy footprint. As evidenced in this study, the parameter constraints introduced by CHQ directly permit the transformation of dense multiplications into simplified shift and add operations. This architectural consequence reduces energy consumption while maintaining the predictive accuracy required by demanding applications. Supported by an automated design flow that unifies model training, quantization, and hardware generation, the constrained-Hamming arithmetic architecture provides an efficient and scalable approach for the future of continuous edge intelligence.

 \bibliographystyle{elsarticle-num} 
 \bibliography{FELIX.bib}





\end{document}